\documentclass{article}

\usepackage{microtype}
\usepackage{graphicx}
\usepackage{subcaption}
\usepackage{booktabs}
\usepackage{multirow}
\usepackage{colortbl}
\usepackage{tikz}
\usepackage{chemfig}
\setchemfig{atom sep=2em}
\usepackage{mod}
\usepackage[colorlinks=true, allcolors=Blue]{hyperref}
\hypersetup{
    colorlinks,
    linkcolor = Blue,
    citecolor = blue,
    urlcolor = BlueViolet
}

\usepackage[accepted]{icml2025}
\usepackage{amsmath}
\usepackage{amssymb}
\usepackage{mathtools}
\usepackage{amsthm}

\usepackage{chemfig}
\setchemfig{atom sep=1.8em}
\usepackage{tikz}
\usetikzlibrary{arrows}
\usetikzlibrary{arrows.meta, calc}

\usepackage[capitalize,noabbrev]{cleveref}

\theoremstyle{plain}

\theoremstyle{definition}

\theoremstyle{remark}

\usepackage[textsize=tiny]{todonotes}

\definecolor{mygray}{gray}{0.9}
\definecolor{myBlue}{rgb}{0.0, 0.2, 0.6}
\definecolor{myGreen}{rgb}{0.1, 0.5, 0.0}
\definecolor{myBrown}{rgb} {0.43, 0.21, 0.1}
\definecolor{myCyan}{rgb}{0.0, 0.58, 0.71}
\definecolor{myBlue2}{rgb}{0.36, 0.57, 0.9}
\definecolor{myRed}{rgb}{0.87, 0.36, 0.51}

\icmltitlerunning{Pathways in Reaction Networks using Integer Linear Programming}

\begin{document}

\twocolumn[
\icmltitle{\textsc{Finding Pathways in Reaction Networks\\ guided by Energy Barriers\\ using Integer Linear Programming}}

\icmlsetsymbol{equal}{*}

\begin{icmlauthorlist}
\icmlauthor{Adittya Pal}{sdu,equal}
\icmlauthor{Rolf Fagerberg}{sdu}
\icmlauthor{Jakob Lykke Andersen}{sdu}
\icmlauthor{Peter Dittrich}{fsu}
\icmlauthor{Daniel Merkle}{sdu,bu}
\end{icmlauthorlist}

\icmlaffiliation{sdu}{Department of Mathematics and Computer Science, University of Southern Denmark, Campusvej 55, 5230 Odense M, Denmark\\}
\icmlaffiliation{fsu}{Department of Mathematics and Computer Science, Friedrich Schiller University Jena, F{\"u}rstengraben, 07743, Jena, Germany\\}
\icmlaffiliation{bu}{Faculty of Technology, Bielefeld University, Postfach 10 01 31, 33501 Bielefeld, Germany}

\icmlcorrespondingauthor{Adittya Pal}{adpal@imada.sdu.dk}

\icmlkeywords{chemical reaction network, pathway optimization, integer linear programming, automated search}

\vskip 0.3in
]

\printAffiliationsAndNotice{\icmlEqualContribution} 

\begin{abstract}
Analyzing synthesis pathways for target molecules in a chemical reaction network annotated with information on the kinetics of individual reactions is an area of active study. This work presents a computational methodology for searching for pathways in reaction networks which is based on integer linear programming and the modeling of reaction networks by directed hypergraphs. Often multiple pathways fit the given search criteria. To rank them, we develop an objective function based on physical arguments maximizing the probability of the pathway. We furthermore develop an automated pipeline to estimate the energy barriers of individual reactions in reaction networks. Combined, the methodology facilitates flexible and kinetically informed pathway investigations on large reaction networks by computational means, even for networks coming without kinetic annotation, such as those created via generative approaches for expanding molecular spaces. To demonstrate the methodology, we apply it on a chemical reaction network generated from 2-hydroxyethanenitrile (glycolonitrile), water, and ammonia, where we search for pathways to glycine and 2-hydroxyethanoic acid (glycolic acid) using the input molecules as precursors.   
\end{abstract}

\section{Introduction}\label{sec: introduction}

Kinetic studies of individual reactions have provided valuable insight into their mechanisms, which in turn can generate ways to optimize the reaction to suit the overall synthesis goals.
However, molecules are often entangled in various concurrent reactions, which together form an interconnected web of reactions called a \emph{reaction network}.
Theoretical and computational modeling offers a possibility for analyzing these systems to various degrees of accuracy and computational efficiency, depending on the level of abstraction employed.
In this work, we propose a method for integrating kinetics into the search for pathways in reaction networks modeled on a level of abstraction where the network is represented by a hypergraph and individual reactions are represented by directed hyperedges \cite{jakob_integer_hyperflows}.
The nodes of the hypergraph are molecular graphs (with their nodes representing atoms and edges representing bonds) and hence the modeling is atomistically explicit.
The work is a continuation of our previous work \cite{thermoflow}, which focused on finding pathways in reaction networks based on the thermodynamics of individual reactions in the network. However, despite the name, thermodynamics does not help to elucidate the `dynamics' of the system: how fast does the pathway in question run and how quickly does the system equilibrate? For example, under atmospheric conditions, the conversion of diamond to graphite is thermodynamically favorable. However, the energy barrier for this conversion is high, so, thankfully, for all practical purposes `diamonds are forever'. 
Therefore, kinetic parameters of reactions, such as the free energy barrier or the rate constant, might provide better alternatives for characterizing pathways. 

However, before one can use kinetic parameters to search for pathways in a reaction network, the network itself has to be constructed. This can be done in various ways. In one approach, the potential energy surface of a set of molecules is explored to elucidate trajectories with low-lying saddle points, separating the reactants from the products. Computationally intensive quantum mechanical calculations are used in most cases \cite{crn_expand_qm3}, along with faster but less accurate semi-empirical methods in some \cite{yaks,empi1}.
Recently, generative diffusion models have also been used to perturb molecules to generate reaction networks with unknown mechanisms \cite{OA-ReactDiff}.
A different approach is to use reaction templates for iteratively expanding the network from a set of starting molecules, possibly in conjunction with filters to exclude specific reactions \cite{rmg2}.
Further approaches include the generation of the reaction network of interest in a one-shot heuristical process by an expert~\cite{yaks}, or the loading of an existing network from a database.

Except for the approach where the reaction network is generated by the exploration of the potential energy surface, individual reactions will need to be annotated with the kinetic parameters from a separate source, such as databases \cite{database_energy_barriers}, estimations informed by experiments \cite{reported_network2}, or dedicated quantum mechanical calculations \cite{CRN_qm}.
Empirical relations, deduced via machine learning on available data on kinetics of reactions, have been used to estimate kinetic parameters for reactions \cite{ml_energy_barrier3,ml_reaction_barrier7}. However, these methods have limited success on `out of class' molecules \cite{scaling_limitations}. 

Once a reaction network has been constructed and the reactions annotated with appropriate kinetic data, one can employ diverse strategies to search and analyze pathways in it \cite{search_crn}. Previous work include Beam search based on suitable crafted scoring functions utilizing priority queues \cite{beam_search_pq}, as well as Monte-Carlo search trees extended with a method to sample efficient pathways \cite{mc_trees}.
Traditional path-finding algorithms such as A$^*$ algorithm~\cite{shortest_path_hypergraph2} and cutting-plan algorithm for an integer linear program~\cite{shortest_path_hypergraph1} have been used in previous work to find pathways between molecules.
Some analytical strategies have been used in the study of dynamics of pathways in reaction networks \cite{crn_dynamics2}, but they are often limited by the availability of kinetic data and the scale of the networks.

In this work, we develop an objective measure to rank pathways in a reaction network maximizing the probability of the entire pathway based on physical principles. 
Additionally, based on the notion that energy barriers of reactions are the determining factor for the kinetics of the network, we develop an automated methodology for computing such energy barriers, building on a combination of an array of existing computational tools. A key novelty is that in contrast to the standard manual intervention required to generate reaction trajectories from which the reaction barrier can be estimated, we try to automate this part of the process.
Lastly, we formulate an integer linear program which allows the user to specify queries for pathways in the network in a very flexible and generic way. Solving this integer linear program returns all pathways fulfilling the criteria, ranked by the objective measure mentioned above.
In short, this gives an automated, general, and flexible methodology for analyzing reaction networks, a methodology which offers a speed-up over manual search for pathways but with comparable quality.
To exemplify its utility, we apply it to a reaction network expanded from the molecules 2-hydroxyethanenitrile, ammonia and water, where we query for pathways to glycine, an amino acid, and 2-hydroxyethanoic acid, a molecule encountered in photorespiration in plants.

\section{Definitions}
\label{sec: definitions}

In this section, we present the mathematical frameworks used to model reaction networks and to specify searches for pathways in them,
namely directed hypergraphs and integer hyperflows.
This modeling is based on~\cite{jakob_integer_hyperflows}.

\subsection{Directed Hypergraphs}
\label{sec:hypergraphs}

A reaction network consists of a set of molecules, $V$, and a set of reactions, $E$. Each reaction $e\in E$ generally involves multiple molecules and can be expressed as follows
\begin{equation}
    \sum_{v\in V} s_{ve}^{-} v \longrightarrow \sum_{v\in V} s_{ve}^{+} v ,
    \label{chemical_equation}
\end{equation}
where $s_{ve}^{-}$ and $s_{ve}^{+}$ are stoichiometric coefficients indicating the number of molecules of $v$ consumed and produced, respectively. 

A reaction network can be effectively modeled as a directed multi-hypergraph~\cite{jakob_integer_hyperflows}, denoted as $\mathcal{H} = (V, E)$, where vertices $v \in V$ represent molecules and hyperedges $e \in E$ represent reactions. Each hyperedge $e \equiv (e^-, e^+)$ consists of two bags (multi-sets) of molecules: $e^-$ for reactant molecules and $e^+$ for product molecules. The count of a molecule $v$ in the reactant bag and in the product bag for a reaction $e$ is given by $s^-_{ve}$ and $s^+_{ve}$, respectively. \footnote{This notation differs from~\cite{jakob_integer_hyperflows} by swapping $^+$ and $^-$ as the superscript on $e$.}

\begin{figure*}[t]
    \centering
    \begin{subfigure}{0.4\textwidth}
        \centering
        \includegraphics[width=\textwidth]{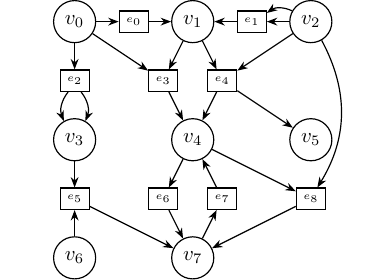}
        \caption{A directed hypergraph (top-left), where circles are vertices and the squares are hyperedges.}
        \label{fig:hypergraph}
    \end{subfigure}%
    \hspace{1cm}
    \begin{subfigure}{0.4\textwidth}
        \centering
        \includegraphics[width=\textwidth]{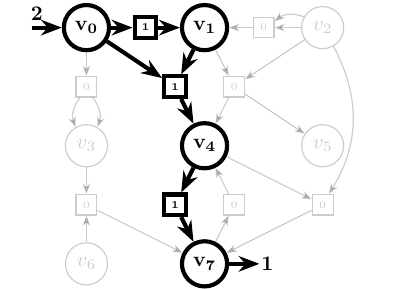}
        \caption{A possible integer hyperflow in the hypergraph, representing a pathway to the queried molecules $v_7$ with $v_0$ as inflow.}
        \label{fig:hyperflow1}
    \end{subfigure}
    \\
    \vspace{1.0cm}
    \begin{subfigure}{0.4\textwidth}
        \centering
        \includegraphics[width=\textwidth]{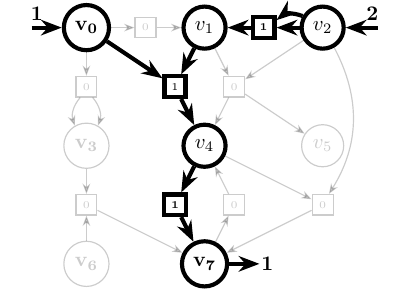}
        \caption{Another integer hyperflow in the hypergraph to the queried molecules $v_7$ with $v_0$ and $v_2$ as inflows.}
        \label{fig:hyperflow2}
    \end{subfigure}%
    \hspace{1cm}
    \begin{subfigure}{0.4\textwidth}
        \centering
        \includegraphics[width=\textwidth]{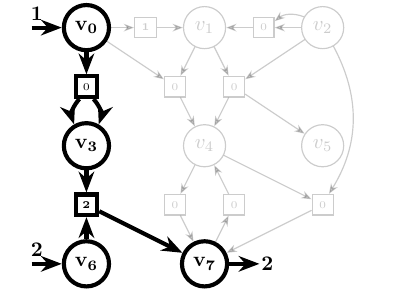}
        \caption{A third integer hyperflow in the hypergraph to the queried molecules $v_7$ with $v_0$ and $v_6$ as inflows.}
        \label{fig:hyperflow3}
    \end{subfigure}
    \caption{
    A multi-directed hypergraph representing a reaction network with three possible integer hyperflows, where hyperedges with non-zero flow are drawn in bold.
    The inflow and the outflow are depicted as arrows in and out of the source and target vertices of the pathway.
    Hyperedges with multiple copies of a vertex as source or target are depicted with parallel arrows.
    See for instance~$e_2$, which has $v_3$ twice as a target and thus represents the reaction $v_0\rightarrow 2\, v_3$.}
    \label{fig: hypergraph_example}
\end{figure*}

Figure~\ref{fig:hypergraph} illustrates a hypergraph with eight vertices and nine hyperedges. It has the one-to-one reaction $e_0 \equiv (\{v_0\}, \{v_1\})$, the many-to-one reactions $e_3 \equiv (\{v_0, v_1\}, \{v_4\})$ and $e_1 \equiv (\{v_2, v_2\}, \{v_1\})$, the one-to-many reaction $e_2 \equiv (\{v_0\}, \{v_3, v_3\})$, and the many-to-many reaction $e_4 \equiv (\{v_1, v_2\}, \{v_4, v_5\})$.

\subsection{Integer Hyperflows}
\label{sec:hyperflows}

A pathway is a sequence of reactions that converts source molecules into target molecules. To formalize pathways, \cite{jakob_integer_hyperflows} introduced the concept of an \emph{integer hyperflow}~$f$ in a hypergraph~$\mathcal{H} = (V,E)$.

An integer hyperflow assigns a non-negative integer to each edge $e \in E$ which specifies the number of times that the reaction represented by~$e$ is used in the pathway.
A pathway effects an overall reaction with a set of source molecules and and a set of target molecules. To model these sets of source and target molecules, 
\cite{jakob_integer_hyperflows} defines auxiliary half-edges for each $v \in V$: $e_v^+ = (\emptyset, v)$ specifies the role of $v$ as a source molecule and $e_v^- = (v, \emptyset)$ specifies the role of $v$ as a target molecule.
Setting
$$
\begin{array}{lcl}
E^+ & = & \{ e_v^+ \mid  v \in V\},\\
E^- & = & \{ e_v^- \mid  v \in V\},\\
\overline{E} & = & E \cup E^+ \cup E^-,\\
\end{array}
$$
the hypergraph is extended to $\overline{\mathcal{H}} = (V, \overline{E})$ and an \emph{integer hyperflow} is defined as a vector $f \in \mathbb{N}_{0}^{|\overline{E}|}$ satisfying the flow conservation constraint:
\begin{equation}
    \forall v\in V\colon \sum_{ e\in \overline{E} } s_{ve}^+ f_e - \sum_{ e\in \overline{E} } s_{ve}^- f_e = 0,
    \label{flow_conservation}
\end{equation}
where $s_{ve}^-$ and $s_{ve}^+$ are the stoichiometric coefficients from Section~\ref{sec:hypergraphs}. Equation \eqref{flow_conservation} is a restatement of the mass conversation: it implies that for all molecules $v\in V$ in the network, the amount of $v$ produced by reactions $e\in E$ where it as a product and supplied by an inflow (by $e_v^+$) must equal the amount consumed by reaction $e\in E$ where it is a reactant and removed by an outflow (by $e_v^-$). For reactions $e$ where molecule $v$ does not appear as a product $s_{ve}^+ = 0$ and for reactions $e$ where molecule $v$ is not used as a reactant $s_{ve}^-=0$, allowing us to extend the sum over all hyperedges $e\in\overline{E}$. A hyperflow $f$ can be considered as an assignment of an integer $(\geq 0)$ to all hyperedge $e\in E$, indicating the number of times the corresponding reaction has to occur to carry out the desired transformation in the pathway.
Integer hyperflows are a formalization of the standard notion and depiction of pathways, which consists of reactions and their integer frequencies, which denotes the number of times that reaction is used in the pathway.

Figure~\ref{fig:hyperflow1}, \ref{fig:hyperflow2} and \ref{fig:hyperflow3} illustrates three example hyperflows in a hypergraph, with unboxed numbers showing flows on half-edges and boxed numbers indicating flows on $E$. Only half-edges with non-zero flow are displayed. As seen from the Figures~\ref{fig:hyperflow1}, \ref{fig:hyperflow2} and \ref{fig:hyperflow3}, integer hyperflows correspond well to the standard notion and depiction of pathways, as these use each reaction an integral number of times in the pathway.

Integer hyperflows offer a versatile way to define pathways by constraining flow on half-edges. As examples, one can specify an overall reaction by fixing flows for source and target molecules, or one can set upper and lower bounds for these flows while allowing specified sets of by-products (potentially also with bounds on their outflow). Overall, this modeling allows the problem of searching for pathways in reaction networks to be transformed into the problem of finding a flow vector~$f$ satisfying Equation~\eqref{flow_conservation} plus a number of specified constraints. Given the integer nature of hyperflows, the language of Integer Linear Programming (ILP) is a natural framework for expressing this integer hyperflow modeling, which allows the search for the specified pathways to be carried out by any of the many ILP-solvers available. This approach is part of what the software library MØD~\cite{mod_web} implements.

A discussion on how integer hyperflows relate to and differ from similar methods such as flux balance can be found in Section 2.6 of~\cite{jakob_integer_hyperflows}. Moreover, an alternative exposition, aiming to be less formal, of how reaction networks can be represented as hypergraphs and how pathways in them can be represented as integer hyperflows can be found in Appendix~\ref{first appendix}.

\section{Methods}
\label{sec: methods}

In this section, we detail our proposed methodology for automating the search for kinetically informed pathways in reaction networks. In Sections~\ref{reaction_prob} and \ref{obj_func}, we first propose an objective function for ranking pathways in a reaction network annotated with kinetic data in the form of energy barriers for each reaction. In Section~\ref{implemented_model}, we then formulate an integer linear program (ILP) representing such networks and allowing the specification and execution of pathway searches. This part builds on the modeling described in~Section~\ref{sec: definitions}. We also describe how to search for not only one, but multiple solutions to a given pathway query, ordered by an objective function such as the one presented in Section~\ref{obj_func}. Lastly in Section \ref{kinetic_oracle}, we describe our computational workflow for estimating the energy barriers of reactions, a workflow which enables the automatic annotation of reaction networks with kinetic data.

\subsection{From Rate Constants to Edge Weights}
\label{reaction_prob}

Irrespective of whether one uses Arrhenius' theory \cite{arrhenius1,arrhenius2}, collision theory or transition state theory \cite{eyring_equation}, the dependence of the rate constant~$k_e$ of a reaction~$e$ on the free energy barrier $G_e$ follows that of a negative exponential. The expression for the rate constant by Eyring's equation from transition state theory is
\begin{equation}
    k_e = \frac{k_b T}{h} \exp\left(-\frac{G_e}{RT}\right) \label{energy_barrier}
\end{equation}
where $R$ is the universal gas constant, $k_b$ is the Boltzmann constant, and $h$ is the Planck constant, each in their appropriate units.
In a well-stirred system where all the reactions in the reaction network are assumed to be concurrently taking place, the temperature can be considered to be the same for all reactions. Hence, it is not a factor affecting the relative rates of the reactions in the system.

Since our modeling is not aimed at capturing temporal changes in the concentrations of molecules, we make the simplifying assumption that all molecules in the reaction network are present with unit concentrations. Under this assumption, the rate constant $k_e$ can be interpreted as being proportional to the probability of reaction~$e$ happening in a fixed small volume and time interval. 
A reaction with a higher rate constant (and lower energy barrier) is expected is occur more frequently, hence the probability of that reaction occurring in a fixed interval of time would be higher.
To interpret the rate constant as probabilities of the reaction in a given reaction network, we normalize these values to sum to one:
\begin{align}
    p(e) &= \frac{k_e}{\sum_{i\in E} k_i} \nonumber \\ 
    &= \frac{\frac{k_b T}{h} \exp\left(-\frac{G_e}{RT}\right)}{\sum_{i\in E} \frac{k_b T}{h} \exp\left(-\frac{G_i}{RT}\right)}\tag{\text{using }\ref{energy_barrier}} \\
    &= \frac{\exp\left(-\frac{G_e}{RT}\right)}{\sum_{i\in E} \exp\left(-\frac{G_i}{RT}\right)} \nonumber \\
    &= \frac{\exp\left(-\frac{G_e}{RT}\right)}{D} \label{prob_reaction}
\end{align}
Here, the denominator $\sum_{i\in E} \exp(-\frac{G_i}{RT})$ is the same for all reactions~$e$ and has been shortened to~$D$. In this paper, we assign the measure~$p(e)$ as the weight of the hyperedge representing~$e$.
However, the user is free to assign any other measures of weight to the hyperedges in the reaction network, if deemed better suited for a particular pathway search objective.

\subsection{Formulating an Objective Function}
\label{obj_func}
We are interested in pathways from the given source molecules to the desired target molecule(s) that have a high overall probability of occurring in practice. Considering individual reactions in a pathway as independent events (and abusing the notation~$e$ to mean the event of reaction~$e$ happening), the probability~$p_{\text{total}}$ of the entire set of reactions in the pathway occurring in one unit of time and one unit of volume is
\begin{align*}
    p_{\text{total}} &= p\left( \bigwedge_{e: f_e > 0} \left(e\wedge e \wedge \ldots f_e\text{ times}\right)\right) \\
    &= \prod_{e: f_e > 0} p(e)^{f_e} \\
    &= \prod_{e: f_e > 0} \left[\frac{\exp\left(-\frac{G_e}{RT}\right)}{D}\right]^{f_e} \tag{\text{using }\ref{prob_reaction}} \\
    &= \frac{\prod_{e: f_e > 0} \exp\left(-\frac{f_e\cdot G_e}{RT}\right)}{\prod_{e: f_e > 0} D^{f_e}} \\
    &= \left.\exp\left(-\frac{\sum_{e \in E} f_e\cdot G_e}{RT} \right) \middle/ D^{\sum_{e \in E} f_e}\right.
\end{align*}
Since the energy barrier $G_e$ is weighted by the flow $f_e$, the conditional sum $e: f_e > 0$ can be substituted by a simple sum over all hyperedges $e\in E$ because hyperedges with $f_e = 0$ do not contribute to the value of the sum.
The expression for the total probability of a pathway is a fraction and we take the logarithm of it to simplify it and make it a linear expression.
\begin{align*}
    \log p_{\text{total}} &= \log \left[\exp\left(-\frac{\sum_{e \in E} f_e\cdot G_e}{RT} \right) \middle/ D^{\sum_{e \in E} f_e}\right] \\
    &= -\frac{\sum_{e \in E} f_e\cdot G_e}{RT} - \left(\sum_{e \in E} f_e\right) \cdot \log D 
\end{align*}
The last line is proportional to the following expression obtained by multiplying by $RT$
\begin{equation*}
    -\sum_{e \in E} f_e\cdot G_e - RT\cdot \log D\cdot \sum_{e \in E} f_e
\end{equation*}
The probability of the set of reactions in the pathway is then maximized by maximizing the value of the obtained expression:
\begin{equation*}
    \max\left(-\sum_{e \in E} f_e\cdot \left(G_e + RT\cdot \log D \right)\right)
\end{equation*}
This is equivalent to the following minimization problem:
\begin{align}
    \min\left(\sum_{e \in E} f_e\cdot \left(G_e + RT\cdot \log D \right)\right) \label{proposed_objective}
\end{align}
This proposed objective value~\eqref{proposed_objective} is a linear expression because the $G_e$'s are constants from the input instance, $T$ is assumed constant, and $R$ is a physical constant. Also, the value $\log D = \log \left(\sum_{i\in E} \exp\left(-\frac{G_i}{RT}\right)\right)$ is constant when considering a fixed reaction network.

We note that via $D$, the objective value~\eqref{proposed_objective} of a pathway will depend on the underlying reaction network in which it is considered embedded. This may be regarded as a weakness. On the other hand, the objective value \eqref{proposed_objective} is intuitively appealing by being composed of two natural components: Its first component $\sum_{e \in E} f_e\cdot G_e$ favors pathways with small sums of energy barriers of the reactions in the pathway, weighted by the number of times it is used. Its second component $\sum_{e \in E} f_e \cdot \left( RT \, \log D\right)$ favors pathways with fewer reactions. The factor $ RT \log D$ can be viewed as weighting the relative importance of the two components.

\subsection{ILP Modeling} \label{implemented_model}

In this section, we express the modeling of Section~\ref{sec: definitions} as an integer linear program (ILP). Recall the key idea that in a reaction network, pathways can be expressed as hyperflows and pathway queries can be expressed as constraints on hyperflows. Such constraints can simply be added to the ILP, after which the pathway queries can be executed by running an ILP-solver on the final ILP. We also describe how to search for multiple, structurally different solutions to a given pathway query, ordered by an objective function such as~\eqref{proposed_objective}.

Thus, in the ILP a pathway is represented by a vector of non-negative\footnote{Non-negative flow values require reversible reactions to be modeled as two separate hyperedges in the reaction network, but it also allows the flexibility to model some reactions as irreversible.} integer-valued variables~$f_e$, one variable for each reaction~$e$ in the reaction network. These flow variables are subject to the key constraint~\eqref{flow_conservation}, as well as to the user-supplied constraints that specify the pathway query under consideration.

To model reversible reactions, separate hyperedges were added for the forward and backward directions of the reaction. This allows us to restrict the flows to non-negative integers only. The use of the reverse of a reaction would be denoted by a positive flow on the hyperedge in the opposite direction.
\begin{equation}
    f_e \geq 0 \label{flow_bounds}
\end{equation}

For each reaction~$e$, we add a boolean indicator variable $z_e$ which keeps track of whether the hyperedge~$e$ is used in the pathway or not. The values of these variables are assigned according to their respective flow variables following the biimplication constraints:
\begin{align}
    f_e = 0 &\iff z_e = 0 \label{biimp1} \\
    f_e \geq 1 &\iff z_e = 1 \label{biimp2}
\end{align}
We will use these indicator variables later when listing multiple, structurally different solutions to a given pathway query.\footnote{The indicator variables can also increase expressiveness when specifying pathway queries.} Many ILP solvers support addition of such indicator constraints \eqref{biimp1} and \eqref{biimp2} directly as implications to an ILP problem (which are then solved by a nonlinear, nonconvex reformulation of the indicator constraint). If this feature is not available, one can instead use the method of linearizing these implication constraints via a large constant~$M$.

\begin{table}[t]
    \centering
    \begin{tabular}{ccr}
        \toprule
        Constant & Datatype & Description \\
        \midrule
        $G_e$ & float & Weight of the hyperedge $e$ \\[1mm]
        $s_{ve}^+$ & integer & Count of molecule $v$ \\
        && on the LHS of a reaction $e$ \\[1mm]
        $s_{ve}^-$ & integer & Count of molecule $v$ \\
        && on the RHS of a reaction $e$ \\[1mm]
        $RT$ & float & Product of \\
        && universal gas constant \\
        && and temperature \\[1mm]
        $M$ & integer & Big integer used for \\
        && linearizing the constraints \\
        && (type promoted to float,\\
        && if required) \\
        \bottomrule
    \end{tabular}
    \caption{Constants in the proposed ILP formulation.}
    \label{tab: ILP constants}
\end{table}%

In summary, the resulting ILP looks as follows:

\begin{align}
    & \min\left(\sum_{e \in E} \left(f_e\cdot G_e + RT\cdot \log D\cdot f_e\right)\right) \tag{\ref{proposed_objective}} \\
    \forall e \in E\colon & 0 \leq f_e \tag{\ref{flow_bounds}} \\
    \forall v\in V \colon &\sum_{ e\in \overline{E} } s_{ve}^+ f_e - \sum_{ e\in \overline{E} } s_{ve}^- f_e = 0 \tag{\ref{flow_conservation}}\\
    &\text{Constraints specifying the pathway query} \nonumber\\
    & f_e = 0 \iff z_e = 0 \tag{\ref{biimp1}} \\
    & f_e \geq 1 \iff z_e = 1 \tag{\ref{biimp2}}
\end{align}

The constraints specifying the pathway consist of the desired outflow of the queried target molecule and the supplied source molecules. Additionally, it might involve constraints such an disallowing particular molecules as by-products (by restricting their outflow to zero), or disallowing the use of particular reactions (by specifying $f_e = 0 = z_e$ for those reactions). An overview of the variables and constants in the ILP can be found in Table \ref{tab: ILP variables} and \ref{tab: ILP constants}, respectively.

\begin{table}[t]
    \centering
    \begin{tabular}{ccr}
        \toprule
        ILP variable & Datatype & Description \\
        \midrule
        $f_e$ & integer & Flow variable for\\
        && a hyperedge $e$ \\
        && non-negative \\[1mm]
        $z_e$ & boolean & Indicator variable \\
        && denoting if a hyperedge \\
        && $e$ is used in the flow \\
        \bottomrule
    \end{tabular}
    \caption{Variables in the proposed ILP formulation.}
    \label{tab: ILP variables}
\end{table}%

\subsubsection{Enumeration of solutions}

\begin{figure}[b]
    \centering
    \includegraphics[scale=1.0]{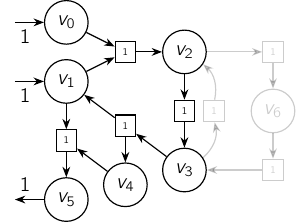}
    \caption{An example of a solution (the hyperflow of the entire figure) which contains a previous solution (the hyperflow shown in black) as a subset, seen in terms of the edges (reactions) used. Addition of constrain~\eqref{subset_elimination} for all previously returned solutions ensures that such solutions are not returned.} 
    \label{fig: cover}
\end{figure}

We would like to enumerate multiple solutions to a pathway query in the reaction network. In this enumeration, we prefer to get a diverse set of pathways, in order to make the result more robust to uncertainties in the weights (i.e., the energy barrier values) assigned to the hyperedges. 
Such diversity offers a selection of alternative pathways for implementation, which is beneficial in case the single optimal solution turns out to be unattainable in practice (or just less desirable than expected due to circumstances not included in the modeling).
Some ILP solvers provide so-called solution pools which allow enumeration of solutions, but the process of populating this pool of solutions is often opaque.
More control is given in M{\O}D~\cite{mod_web}, which provides an option to enumerate multiple solutions based on a user-supplied specification of which subset of integer and binary variables to enumerate over~\cite{jakob_integer_hyperflows}.
For example, if the specification lists exactly the variables $z_e$ for each $e\in E$, then two pathways using the same set of hyperedges (but potentially with a difference in the flow value on some edges) will be considered equal,
and only the one with the best objective value is returned.
That is, if we define $supp(f) = \{e\in E \mid z_e = 1\}$ then two solutions $f_1$ and  $f_2$ are considered different exactly when $supp(f_1) \neq supp(f_2)$.

However, in our case initial experimentation showed that using this enumeration policy, later enumerated solutions $f_2$ often contained a previous solution $f_1$, in the sense that $supp(f_1) \subsetneq supp(f_2)$, as exemplified in Figure~\ref{fig: cover}.
To increase diversity of solutions, we opted for a custom enumeration scheme where $supp(f_2)$ of new solution~$f_2$ cannot contain $supp(f_1)$ of a previous solution~$f_1$ as a subset. When $supp(f_1) \subsetneq supp(f_2)$ holds for for a pair of solutions, the following statement must evaluate to \textsc{true} for the solution $f_1$:

\begin{center}
\begin{tabular}{lrl}
    & $\neg \bigwedge_{e\in supp(f_2)} z_e$ & $=$ TRUE\\
    or,& $\bigvee_{e\in supp(f_2)} \neg z_e$ & $=$ TRUE\\
    or,& $\sum_{e\in supp(f_2)} (1 - z_e)$ & $\geq 1$\\
    or,& $|supp(f_2)| - \sum_{e\in supp(f_2)} z_e$ & $\geq 1$\\
    or,& $\sum_{e\in supp(f_2)} z_e$ & $\leq |supp(f_2)| - 1$
\end{tabular}
\end{center}

In more detail: Let $f_i$ be the $i$th solution found. 
We then disallow subsequent solutions to contain $supp(f_i)$ by requiring that at least one of the hyperedges of~$f_i$ is not used. This is expressed via the following constraint:
\begin{equation}
    \sum_{e\in supp(f_i)} z_e \leq |supp(f_i)| - 1
    \label{subset_elimination}
\end{equation}
This constraint ensure that not all of the hyperedges in solution $f_i$ is used in some subsequent solution.
When the $k$th solution is to be found, there has thus been added $k-1$ of these constraints in total.
However, this method works only when the set of variables enumerated over is boolean, as is the case with $z_e$.

\subsection{Computational Framework}
\label{kinetic_oracle}

\begin{figure}
    \centering
    \includegraphics[width=\linewidth]{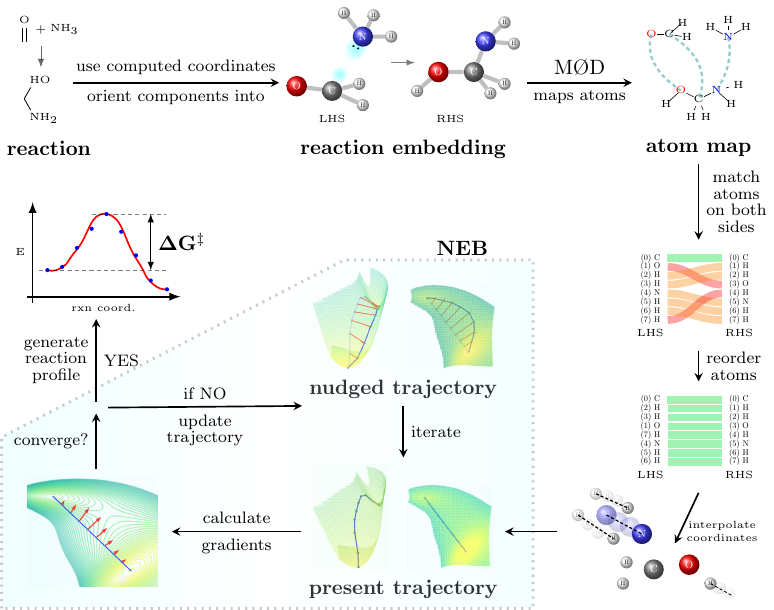}
    \caption{Scheme depicting the workflow used to estimate the energy barrier ($G_e$) for each reaction in the network, represented by a hyperedge.}
    \label{fig:workflow2}
\end{figure}

Above in Section \ref{implemented_model}, we have described our method for searching for pathways in a reaction network annotated with energy barriers for each reaction. In this section, we develop a computational workflow for estimating the energy barriers of reactions, a workflow aimed at automatic annotation of reaction networks with kinetic data. An overview of the workflow to assign the geometries to the molecular graphs in the vertices of the hypergraph representation of the reaction network is given in Figure \ref{fig:workflow1}, and to annotate the hyperedges with the energy barrier for the corresponding reactions represented is given in Figure \ref{fig:workflow2}.

\subsubsection{Method for Estimating Geometry of the Molecules}

We represent individual molecules in the reaction network as molecular graphs. Vertices represent atoms and are labeled by their element name, while edges represent bonds and are labeled by their bond type (single, double, triple or aromatic bond).
Most tools for estimating the energy barriers of reactions require the three-dimensional embedding of the involved molecules with coordinates for each atom. We generate these molecular embeddings using Open Babel and then refine it using xTB in the following manner:

\begin{figure}
    \centering
    \includegraphics[width=\linewidth]{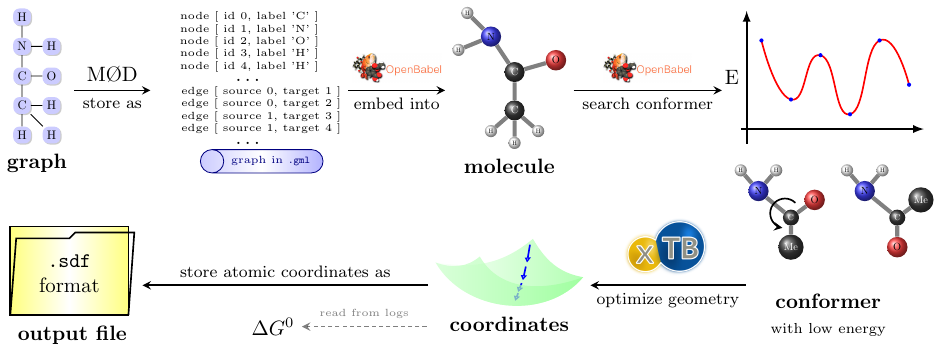}
    \caption{Scheme depicting the workflow used for the assignment of geometries to the molecular graphs in the nodes of the hypergraph.}
    \label{fig:workflow1}
\end{figure}

\begin{itemize}
    \item The molecular graph is loaded into Open Babel as a molecule object.
    \item An initial guess for the three-dimensional embedding of the molecule is generated based on a combination of elementary rules and frequently encountered fragments by using the \texttt{make3D()} \cite{openBabel_make3d} function of Open Babel. However, often these guessed structures can have substructures with steric clashes or high strain in them and therefore the structures need to be optimized.
    \item A systematic search for a lower energy conformer is performed by altering the torsional angles and estimating the energy of the resulting conformer using an elementary force field method (the universal force field (UFF) along with the \texttt{SystematicRotorSearch()} \cite{openBabel_systematicRotorSearch} function of Open Babel. 
    \item The coordinates of each atom in this resulting low energy conformer is written out in a format which can be fed as input to xTB.
    \item Optimization of the generated geometry is performed by xTB \cite{xtb_geomOpt} until the structure generated by its gradient descent based on the forces on individual atoms has converged. 
    \item A single-point calculation can be performed on this optimized structure to find the Gibbs free energy of the molecules. This step is optional and can be performed if the chemical potentials of the molecules are of interest in the modeling of the reaction network.   
\end{itemize}

\subsubsection{Method for Estimating Energy Barriers for the Reactions}

The pipeline described above generates the coordinates for the atoms in an individual molecule. However, this data cannot be directly used to determine the energy barriers for reactions involving those molecules because one needs to embed the molecules with proper orientations to form a complex (a molecular graph with multiple connected components). Furthermore, the embedding of the complexes on both sides of the reaction have to be aligned such that interpolated atomic coordinates in intermediate configurations do not pass through an artificial high-energy arrangement. We implement the following steps for calculating the energy barriers:

\begin{itemize}
    \item The coordinates of the molecules on each side of a reaction are read by RDKit to create a collection of molecules. A complex containing these molecules is generated by aligning entire molecules so that the reactive centers are close to each other and in proper spatial orientation. The geometries of individual connected components are left unchanged. 
    \item The embedding is refined by xTB to take into account the inter-molecular interactions, using semi-empirical methods. (For reactions with a single molecule on one of the sides, the optimized structure from the previous section is used directly for that molecule, skipping this step and the previous step.)
    \item For the energy barrier for the reaction to be estimated correctly, the order of the atoms in files containing the coordinates of the atoms for the endpoints must be the same. Our method of generating the reaction network (see Section~\ref{sec: results}) allows us to map corresponding atoms when applying a reaction template to a substructure match in a molecular graph \cite{atom_tracking_mod}. This atom map is used to number the atoms on the two sides of the reaction correspondingly.
    \item The nudged elastic band \cite{neb} method, implemented in the Atomic Simulation Environment \cite{ase-paper}, is used to determine the energy barrier for a reaction \footnote{The nudged elastic band is a double-ended method, which is suitable in our case because both sides of the reaction are known beforehand, in contrast to single-ended methods where only the reactants are known beforehand.}. A number of intermediate images are generated by interpolating the coordinates of the atoms on the two sides of the reaction for an initial trajectory guess.    
    \item In previous work \cite{neuralneb}, a polarizable atom interaction neural network \cite{painn} was trained on a dataset of molecular configurations around reaction trajectories, Transition1x \cite{transition1x}, to estimate the potential energy of the configuration and the forces on the atoms.
    This pre-trained neural net from \cite{neuralneb} we use to estimate the forces on the atoms (given by the gradient of potential) in each intermediate image. By the nudged elastic band method, the computed forces on individual atoms are used to nudge their position to form a lower energy configuration. Iteration of this step lowers the energy of the images, leading to a reaction trajectory with lower energy barriers along its way. When the maximum of the forces on individual atoms falls below a given threshold, the procedure is stopped.
    \item A cubic spline interpolation of the energies of the intermediate images is performed. The energy barrier for the reaction is read off as the difference between the maximum along this interpolation and its starting point.
\end{itemize}

We note that the user of course is free to substitute the above workflow with other procedures for estimating energy barriers for the reactions of the network, or to read values directly from some database, all depending on the desired accuracy of modeling, the available data, and the available computational resources. The crux of our proposed methodology is that it is automated and fairly fast, which makes it applicable in situations where the available data and computational resources are inadequate, such as when using generative approaches for expanding molecular spaces.

\section{Results}\label{sec: results}

In this section, we apply the methods from Section~\ref{sec: methods} to a reaction network generated in silico from 2-hydroxyethanenitrile, ammonia (NH$_3$), and water (H$_2$O), a network previously studied in \cite{implemented_CRN}. The choice of 2-hydroxyethanenitrile as the starting molecule was driven by the fact that it has been reported to occur in interstellar medium as a product of the condensation of methanal with hydrogen cyanide and act as a precursor to important molecules in pre-biotic systems.~\cite{MeCN_precursor} The previous work~\cite{implemented_CRN} constructed a free-energy map of the molecules that might be present when 2-hydroxyethanenitrile reacts with water and ammonia using computational-intensive density functional theory calculations. It focused on finding thermodynamic sinks and kinetic barriers in the different regions of the free energy landscape.
In the current work, the network was expanded using the graph transformation based generative framework M{\O}D \cite{mod_web}. The generation of the network, starting from the three above-mentioned molecules is detailed in Appendix~\ref{second appendix}, through recursive application of the three graph transformation rules from Appendix~\ref{third appendix}.
A graph transformation rules is a template representing a class of reactions, which share the same set of atoms involved in the reaction, with similar changes in the bond set. The left hand side of the rules is a subgraph (representing a molecular fragment), which can be rewritten as the subgraph on the right hand side, when a match is found in one of the graphs representing molecules already present in the reaction network. This executes a reaction, adding the new product molecules and the reaction the reaction network, and expands the molecular space.
 
Molecules with up to two carbon, four nitrogen, and four oxygen atoms were included in the generated network for comparability with \cite{implemented_CRN}, which considered molecules with two carbon atoms only. The constructed reaction network had $44$ vertices and $116$ hyperedges, and shown in Figure~\ref{fig: hypergraph_for_RN}. 
A list of molecules and reactions in the generated network is available in the GitHub data repository.  
To find probable pathways to biologically important molecules of glycine and 2-hydroxyethanoic acid, using 2-hydroxyethanenitrile, NH$_3$, and H$_2$O, an ILP was formulated to query the desired pathways.
The formulated ILP had $336$ variables and $1561$ constraints.

\begin{figure}[thb]
    \centering
    \includegraphics[width=\linewidth]{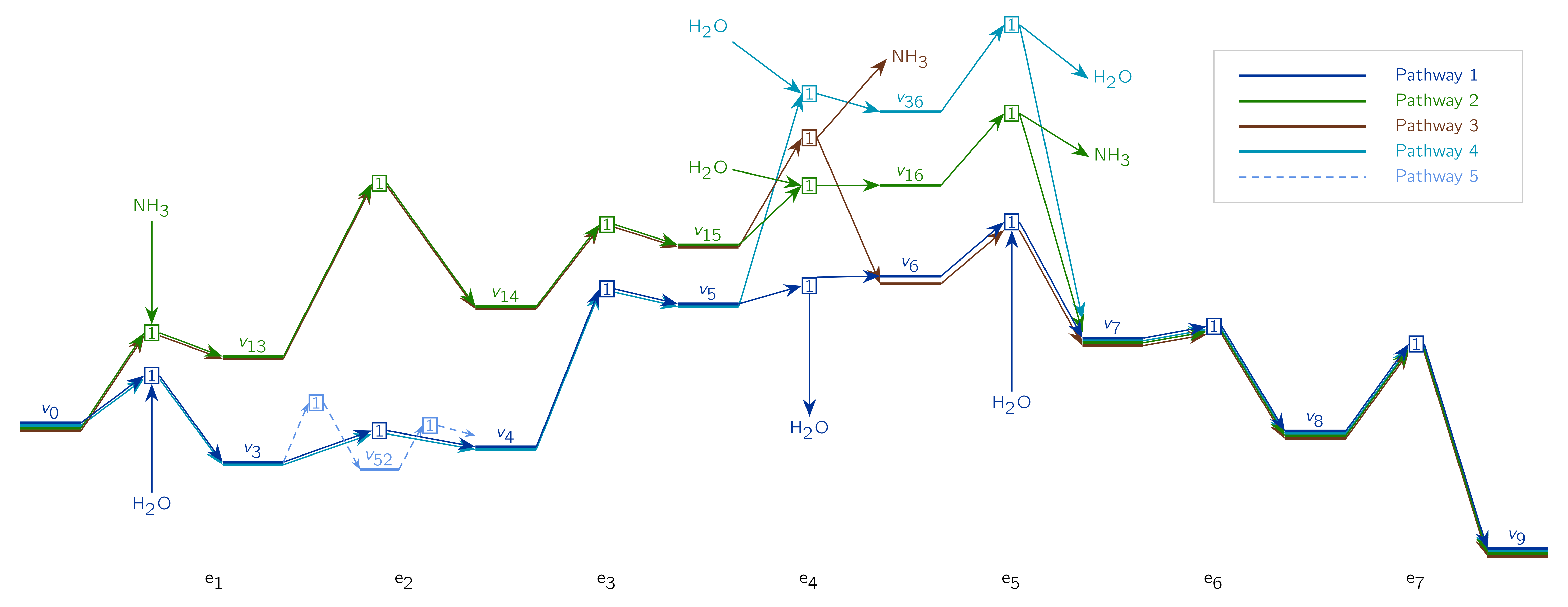}
    \vspace{0.2cm}\\
    \includegraphics[width=\linewidth]{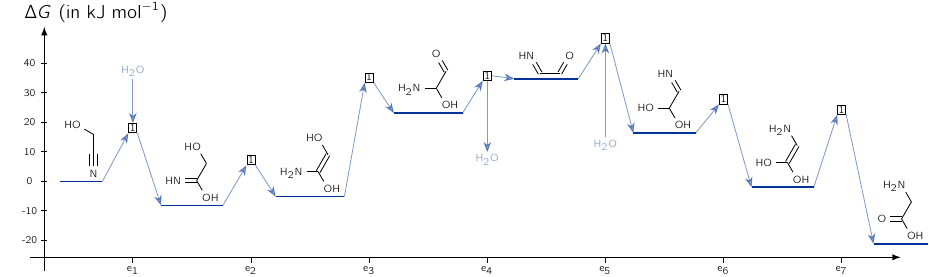}
    \caption{Top: A schematic depiction of the five best pathways to glycine according to the chosen scoring function (energy levels not shown exactly to scale).  
    The structures for the molecules in the corresponding vertices is listed in Table \ref{tab:molecular_structure} and the weights assigned to the hyperedges listed in Table \ref{tab:pathways1}. 
    Bottom: The energy profile of the best-scoring pathway (in blue) from above (energy levels shown to scale).}
    \label{fig:pathway1}
\end{figure}

The five best scoring pathways to the target glycine is depicted in Figure \ref{fig:pathway1}.
The \textcolor{myBlue}{best scoring pathway} scores $143118$ on the chosen objective \eqref{proposed_objective} and proceeds through 2-iminoethanal ($v_6$), a pre-biotic intermediate hypothesized to be of significance. 
A water molecule acts as a catalyst in this pathway since it is used in reaction $e_1$ and generated back in reaction $e_4$. 
The \textcolor{myGreen}{second best pathway} scores $143190$ on the objective \eqref{proposed_objective}. The last two reactions are the same for both (in fact, for all \emph{five}) pathways. 
This second best pathway uses an ammonia molecule as a catalyst, molecules with a higher chemical potential (under unit concentrations) and reactions with higher energy barriers. The sum of the energy barriers for the reactions in this pathway is $200.02$ kJ mol$^{-1}$ as opposed to $128.74$ kJ mol$^{-1}$ for the best pathway. 
One might note that the difference in the objective value of the best and the second-best solutions, that is $143190-143118 = 72$ is entirely due to the difference in the sum of the energy barriers of the reactions along the pathway ($(200 - 128)$ kJ mol$^{-1} = 72$ kJ mol$^{-1}$) or the first part of the ovjective \eqref{proposed_objective} because both the pathways have the same length. Both pathways consist of $6$ reactions and the second part of the objective \eqref{proposed_objective} is the same for both solutions.

\begin{table*}[thb]
    \centering
    \begin{small}
    \begin{tabular}{lrrrrr}
        \toprule
        Reaction\hspace{0.2cm} & \multicolumn{5}{c}{Energy barriers (in kJ mol$^{-1}$)} \\
        count & \hspace{1cm}\textcolor{myBlue}{pathway 1} & \hspace{1cm}\textcolor{myGreen}{pathway 2} & \hspace{1cm}\textcolor{myBrown}{pathway 3} & \hspace{1cm}\textcolor{myCyan}{pathway 4} & \hspace{1cm}\textcolor{myBlue2}{pathway 5} \\
        \midrule
        \multirow{2}{*}{$e_1$} & $13.95$ & $33.18$ & \cellcolor{mygray}$33.18$ & \cellcolor{mygray}$13.95$ & \cellcolor{mygray}$13.95$\\
        & $(v_0+v_2, v_3)$ & $(v_0+v_1, v_{13})$ & \cellcolor{mygray}$(v_0+v_1, v_{13})$ & \cellcolor{mygray}$(v_0+v_2, v_3)$ & \cellcolor{mygray}$(v_0+v_2, v_3)$  \\
        \multirow{2}{*}{$e_2$} & $7.67$ & $64.17$ & \cellcolor{mygray}$64.17$ & \cellcolor{mygray}$7.67$ & $18.67, 13.15$\\
        & $(v_3, v_4)$ & $(v_{13}, v_{14})$ & \cellcolor{mygray}$(v_{13}, v_{14})$ & \cellcolor{mygray}$(v_3, v_4)$ & $(v_3, v_{52}, v_4)$ \\
        \multirow{2}{*}{$e_3$} & $57.69$ & $28.40$ & \cellcolor{mygray}$28.40$ & \cellcolor{mygray}$57.69$ & \cellcolor{mygray}$57.69$\\
        & $(v_4, v_5)$ & $(v_{14}, v_{15})$ & \cellcolor{mygray}$(v_{14}, v_{15})$ & \cellcolor{mygray}$(v_4, v_5)$ & \cellcolor{mygray}$(v_4, v_5)$\\
        \multirow{2}{*}{$e_4$} & $1.67$ & $19.86$ & $29.35$ & $67.36$ & \cellcolor{mygray}$1.67$\\
        & $(v_5, v_6+v_2)$ & $(v_{15}+v_2, v_{16})$ & $(v_{15}, v_6+v_1)$ & $(v_5+v_2, v_{36})$ & \cellcolor{mygray}$(v_5, v_6+v_2)$\\
        \multirow{2}{*}{$e_5$} & $17.14$ & $23.79$ & \cellcolor{mygray}$17.14$ & $60.84$ & \cellcolor{mygray}$17.14$ \\\
        & $(v_6+v_2, v_7)$ & $(v_{16}, v_7+v_1)$ & \cellcolor{mygray}$(v_6+v_2, v_7)$& $(v_{36}, v_7+v_2)$ & \cellcolor{mygray}$(v_6+v_2, v_7)$\\
        \multirow{2}{*}{$e_6$} & $0.19$ & \cellcolor{mygray}$0.19$ & \cellcolor{mygray}$0.19$ & \cellcolor{mygray}$0.19$ & \cellcolor{mygray}$0.19$ \\
        & $(v_7, v_8)$ & \cellcolor{mygray}$(v_7, v_8)$ & \cellcolor{mygray}$(v_7, v_8)$ & \cellcolor{mygray}$(v_7, v_8)$ & \cellcolor{mygray}$(v_7, v_8)$\\
        \multirow{2}{*}{$e_7$} & $30.42$ & \cellcolor{mygray}$30.42$ & \cellcolor{mygray}$30.42$ & \cellcolor{mygray}$30.42$ & \cellcolor{mygray}$30.42$ \\
        & $(v_8, v_9)$ & \cellcolor{mygray}$(v_8, v_9)$ & \cellcolor{mygray}$(v_8, v_9)$ & \cellcolor{mygray}$(v_8, v_9)$ & \cellcolor{mygray}$(v_8, v_9)$\\
        \midrule
        sum & $128.74$ & $200.02$ & $202.85$ & $238.13$ & $152.90$ \\
        \bottomrule
    \end{tabular}
    \end{small}
    \caption{From the left, the columns list the weights assigned to the hyperedges along with the sum of the weights, for the five pathways in Figure \ref{fig:pathway1}.}
    \label{tab:pathways1}
\end{table*}

The \textcolor{myBrown}{third pathway} branches off the second pathway and merges into the first, with a different reaction $e_4$ eliminating an ammonia molecule. 
The \textcolor{myCyan}{fourth pathway} uses a different molecule instead of iminoacetaldehyde, which has a higher chemical potential, hence is ranked lower than the previous three. 
The \textcolor{myBlue2}{fifth pathway} has \emph{eight} reactions, while the previous four had seven reactions. Despite having a lower sum of the energy barriers ($152.90$ kJ mol$^{-1}$), it is ranked lower than the previous three pathways because it is longer.
This exemplifies how the length of the pathways affects its ranking through the second part of the objective \eqref{proposed_objective}.
Figure \ref{fig:pathway1} shows how interconnected the individual different pathways in the network might be, often with parallel branches and cross-connections, which leads to complex behavior of the system.

The estimated energy barriers for the reactions used in the reported pathways in Figure \ref{fig:pathway1} is presented in Table \ref{tab:pathways1}. Table \ref{tab:pathways1} consists of five columns that list the assigned weights for the hyperedges in the pathways shown in Figure \ref{fig:pathway1}. 
The shaded cells indicate hyperedges that have already been used in a previously enumerated pathway. For instance, all five pathways in Figure \ref{fig:pathway1} share the final two reactions. Consequently, the next four pathways reuse the last two reactions of the first pathway, and the last two rows for the corresponding columns are shaded to reflect this overlap.  

Ab initio molecular dynamics has been used in a previous work~\cite{crn_expand_qm3,glycine_pathway} to simulate the Urey-Miller experiment and find possible pathways to glycine in a prebiotic setting. The reported pathways in~\cite{crn_expand_qm3} also contain reactions catalysed by water and ammonia. The reported pathways to glycine in~\cite{crn_expand_qm3} proceeds through a singlet carbene intermediate, while those in~\cite{glycine_pathway} has a molecule with a three-membered ring, oxiranimine, as an intermediate. However, in the current work, there were no reaction templates used which might lead to the formation of a carbene-like molecule or a three-cycle in a molecule in the generated network. Hence, such a pathway was not observed. Carbenes and three-cycles are highly reactive intermediates, and their formation is a high energy process. On the whole, the pathways presented in this work have much lower reported energy barriers than that reported in these two studies.

\begin{figure}[h]
    \centering
    \includegraphics[width=\linewidth]{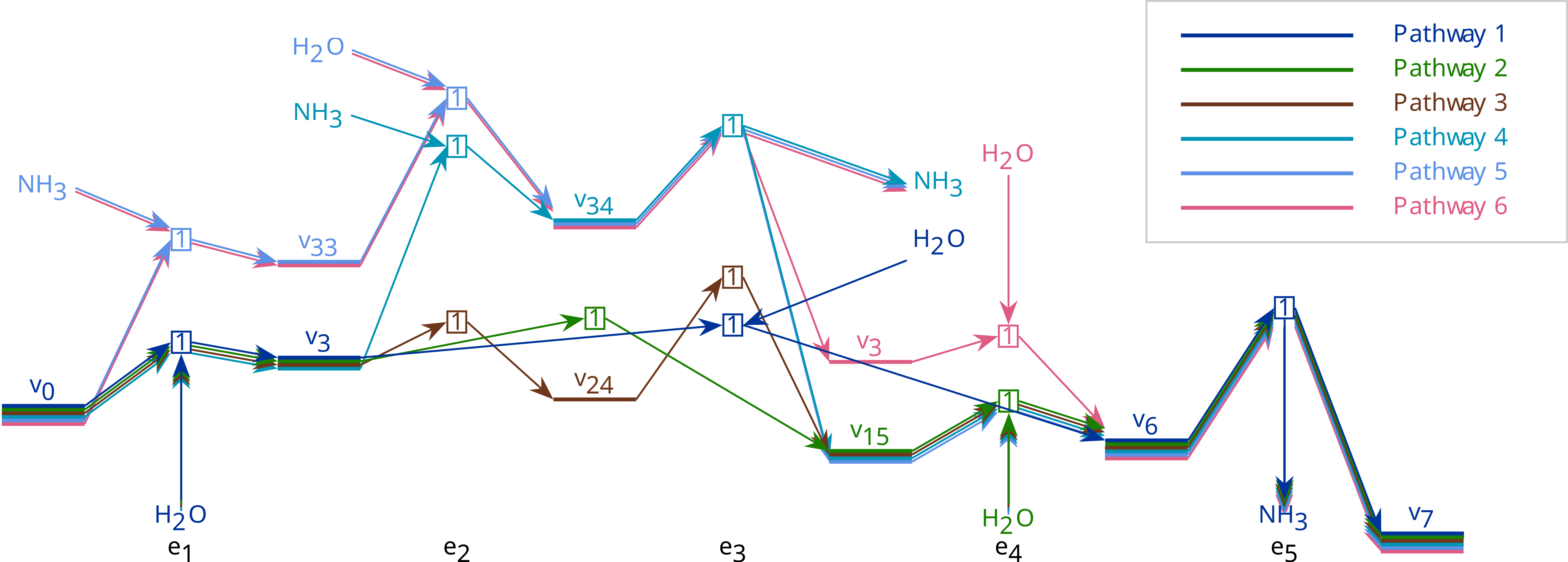}
    \caption{The six best pathways to glyoxlic acid according to the chosen scoring function.
    The structures for the molecules for the vertices is listed in Table \ref{tab:molecular_structure} and the weights assigned to the hyperedges listed in Table \ref{tab:pathways2}.}
    \label{fig:pathway2}
\end{figure}

Another molecule of importance in prebiotic systems which appears in the generated network is 2-hydroxyethanoic acid, commonly known ad glycolic acid. During photorespiration plants produce glycolate. It can also be used as a monomer for various biodegradable polymers such as polyglycolic acid and poly(lactic-co-glycolic) acid. However, there have not been many studies on pathways to glyocolic acid from the precursor molecules in prebiotic systems.  
Figure \ref{fig:pathway2} shows the six best pathways to 2-hydroxyethanoic acid. The \textcolor{myBlue}{best scoring pathway} has three reactions, the \textcolor{myGreen}{second best} has four reaction, while the rest involve five reactions. Despite the third-listed pathway having a lower sum of energy barriers ($108.63$ kJ mol$^{-1}$), it is ranked lower because it is longer, illustrating the effect of the second term in the objective \eqref{proposed_objective}. 

\begin{table*}[t]
    \centering
    \begin{small}
    \begin{tabular}{lrrrrrr}
        \toprule
        Reaction\hspace{0.2cm} & \multicolumn{6}{c}{Energy barrier} \\
        count & \hspace{0.5cm}\textcolor{myBlue}{pathway 1} & \hspace{0.5cm}\textcolor{myGreen}{pathway 2} & \hspace{0.5cm}\textcolor{myBrown}{pathway 3} & \hspace{0.5cm}\textcolor{myCyan}{pathway 4} & \hspace{0.5cm}\textcolor{myBlue2}{pathway 5} & \hspace{0.5cm}\textcolor{myRed}{pathway 6} \\
        \midrule
        \multirow{2}{*}{$e_1$} & $13.95$ & \cellcolor{mygray}$13.95$ & \cellcolor{mygray}$13.95$ & \cellcolor{mygray}$33.18$ & \cellcolor{mygray}$13.95$ & \cellcolor{mygray}$33.18$\\
        & $(v_0+v_2,v_3)$ & \cellcolor{mygray}$(v_0+v_2,v_3)$ & \cellcolor{mygray}$(v_0+v_2,v_3)$ & \cellcolor{mygray}$(v_0+v_1,v_{33})$ & \cellcolor{mygray}$(v_0+v_2,v_3)$ & \cellcolor{mygray}$(v_0+v_1,v_{3}3)$\\
        \multirow{2}{*}{$e_2$} & & & $7.67$ & $48.612$ & $80.53$ & \cellcolor{mygray}$48.61$\\
        & & & $(v_3, v_{24})$ & $(v_{33}+v_2, v_{34})$ & $(v_3+v_1, v_{34})$ & \cellcolor{mygray}$(v_{33}+v_2, v_{34})$\\
        \multirow{2}{*}{$e_3$} & $26.61$ & $18.68$ & $11.76$ & $43.64$ & \cellcolor{mygray}$43.64$ & $33.08$\\
        & $(v_3+v_2, v_6)$ & $(v_3, v_{15})$ & $(v_{24}, v_{15})$ & $(v_{34}, v_{15}+v_1)$ & \cellcolor{mygray}$(v_{34}, v_{15}+v_1)$ & $(v_{34}, v_3+v_1)$\\
        \multirow{2}{*}{$e_4$} & & $0.62$ & \cellcolor{mygray}$0.62$ & \cellcolor{mygray}$0.62$ & \cellcolor{mygray}$0.62$ & \cellcolor{mygray}$0.62$\\
        & & $(v_{15}+v_2,v_6)$ & \cellcolor{mygray}$(v_{15}+v_2,v_6)$ & \cellcolor{mygray}$(v_{15}+v_2,v_6)$ & \cellcolor{mygray}$(v_{15}+v_2,v_6)$ & \cellcolor{mygray}$(v_3+v_2,v_6)$\\
        \multirow{2}{*}{$e_5$} & $74.62$ & \cellcolor{mygray}$74.62$ & \cellcolor{mygray}$74.62$ & \cellcolor{mygray}$74.62$ & \cellcolor{mygray}$74.62$ & \cellcolor{mygray}$74.62$ \\
        & $(v_6, v_7+v_1)$ & \cellcolor{mygray}$(v_6, v_7+v_1)$ & \cellcolor{mygray}$(v_6, v_7+v_1)$ & \cellcolor{mygray}$(v_6, v_7+v_1)$ & \cellcolor{mygray}$(v_6, v_7+v_1)$ & \cellcolor{mygray}$(v_6, v_7+v_1)$\\
        \midrule
        sum & $118.19$ & $107.87$ & $108.63$ & $200.68$ & $213.38$ & $219.11$  \\
        \bottomrule
    \end{tabular}
    \end{small}
    \caption{From the left, the columns list the weights assigned to the hyperedges along with the sum of the weights, for the six pathways in Figure \ref{fig:pathway2}.}
    \label{tab:pathways2}
\end{table*}

Table \ref{tab:pathways2} contains six columns that list the assigned weights for the hyperedges in the pathways illustrated in Figure \ref{fig:pathway2}. As in Table \ref{tab:pathways1}, the shaded cells indicate hyperedges that have already been used in a previously enumerated pathway. The specific hyperedge is mentioned below the energy barrier, with the reactants as the first element of the pair and the products as the second element. Individual molecules are separated by a `+', and the structures of the molecules the labels correspond to can be looked up from Table \ref{tab:molecular_structure}. 

The expansion of the network shown in Figure~\ref{fig: hypergraph_for_RN} took 67 seconds. It is quite faster than comparable methods using reaction templates without using graph transformation, molecular dynamics or potential energy search as done in~\cite{implemented_CRN}. Annotating the $116$ hyperedges with the energy barriers of the reactions as weights took around an hour. Using a neural network instead of more accurate traditional methods does lead to trade off of the accuracy for speed, but it allows scalability for larger networks. Formulating the ILP and enumerating the pathways in the network to the two target molecules using the objective function \eqref{proposed_objective} took around $52$ seconds. All the reported times are the average of $10$ runs on an AMD$^{\circledR}$ Ryzen 5 PRO 4650U CPU (2.1 GHz) utilizing $8$ threads.
One may also expand the network from methanal and hydrogen cyanide instead of 2-hydroxyethanenitrile and use a more general version of the rules where the variable heavy atoms are allowed to match with carbon. nitrogen or oxygen atoms. However, it leads to a larger network with $481$ vertices and $2768$ hyperedges, of which the considered network is a sub-network.

\section{Discussion}\label{sec: discussions}

In this section, we give a comparison of our workflow with existing methods. Constraint-based models such as flux balance analysis also use constraint programming with an objective function that is optimized. For instance, open-source tools such as COBRApy~\cite{cobrapy} perform flux analysis on a reaction network supplied to them. However, such tools do not construct the network on the fly in a generative way--- rather the reaction network has to be constructed beforehand using some other tool. Flux balance analysis generates a flux vector for the reactions in the network which is allowed to contain floating-point values, while our method accepts only integers as a valid hyperflow for a reaction, which is more aligned with the standard notion of a pathway. Formulating a flux balance analysis on the network would be equivalent to an LP relaxation of the pathway query problem using integer hyperflows for the same network. A linear program (LP) can be solved in polynomial time while an ILP is NP-hard. However, formulating the search for pathways to a queried molecule in a reaction network offers some advantages.

\begin{figure}[t]
    \centering
    \begin{subfigure}{0.5\textwidth}
        \centering
        \includegraphics[scale=0.5]{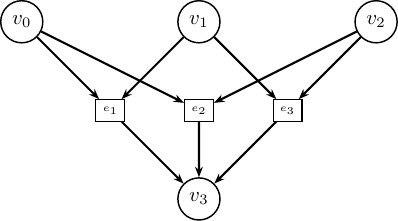}
        \caption{An example reaction network with $4$ molecules and $3$ reactions.}
        \label{fig:example_network}
        \vspace{0.5cm}
    \end{subfigure}
    \begin{subfigure}{0.5\textwidth}
        \centering
        \includegraphics[scale=0.5]{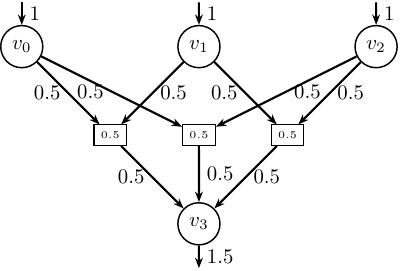}
        \caption{A flux distribution in the network which produces an outflow of $1.5$ for $v_3$.}
        \label{fig:example_flux}
        \vspace{0.5cm}
    \end{subfigure}
    \begin{subfigure}{0.5\textwidth}
        \centering
        \includegraphics[scale=0.5]{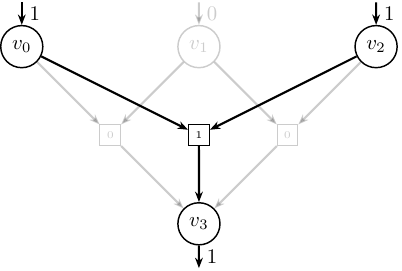}
        \caption{An assignment of integer hyperflows in the network which produces an outflow of $1$ for $v_3$.}
        \label{fig:example_flow}
    \end{subfigure}
    \caption{Comparison of flux balance analysis and an integer hyperflow in an example reaction network.}
    \label{fig:comparison}
\end{figure}

There exists reaction networks where the optimal solutions from the ILP formulated by our modeling and the relaxed LP from flux balance analysis differ. Consider the reaction network shown in Figure~\ref{fig:example_network} with the following formulation of the problem:
\begin{align*}
    &\max \left( \text{outflow}(v_3) \right) \\
    \text{subject to: } &\text{inflow}(v_0) + \text{inflow}(v_1) +\text{inflow}(v_2) \leq 3
\end{align*}
The optimal flux distribution yields an objective value of $1.5$ with the assigned fluxes shown in Figure~\ref{fig:example_flux} while our method assigns the integer hyperflows shown in Figure~\ref{fig:example_flow} in one of the three optimal solutions with objective value $1$. The other two solutions assign the hyperflows $(0,1,1)$ and $(1,1,0)$ to the hyperedge set $(e_1,e_2,e_3)$ respectively.

The ILP solution space is a subset of the LP feasible reaction and therefore, the objective value for the optimal solution for the relaxed LP will always be as at least as large as the objective value of optimal solution for the corresponding ILP. This is illustrated in the example shown in Figure~\ref{fig:comparison} too. As the result, the fluxes and the structure of the solution is often different from those if they were constrained to be integers.

Restricting hyperflows to integers makes physical sense because molecules occur and react in discrete quantities.
Actually, small molecular counts in several biological settings (often involving trace amounts of enzymes and inhibitors affecting multiple iterations of the same pathway) and physical cases (as in chain reactions and cascades) do not allow arbitrary scaling of the number of molecules, as the use of real-valued fluxes by flux balance analysis assume.
When a small number of molecules are encountered in a scenario as above (or in scenarios with appearances of a rare but significant molecule in the system), the discrete nature of the molecules becomes more pronounced.

Another feature of our proposed method is the enumeration of alternative pathways with disjoint set of hyperedges ranked on the basis of a chosen objective function. While flux balance analysis offers the possibility to find alternative pathways~\cite{fba_alternative_pathways}, they have the same objective value but a different flux distribution in the network. This differs from the enumeration of solutions in the current method which offers a list of pathways ordered by the increasing value of the objective function for the user to choose from. Enumerating solutions over variables with a continuous domain is difficult using indicator variables and biimplications as in constraints \ref{biimp1}  and \ref{biimp2}. The enumeration of solutions offers diverse pathways to the queried molecules and makes the method more robust to uncertainties in the weights assigned to the hyperedges.  

Previous work such as~\cite{shortest_path_hypergraph1,shortest_path_hypergraph2} have utilized hyperpaths in directed hypergraphs to depict pathways in reaction networks. However, they use the notion of path length to characterize the goodness of the pathway: a shorter pathway is considered better. Often this might not be the case as the shortest pathway might contain a slow or a high-cost reaction, which becomes a bottleneck for the entire pathway. Therefore, it is important to consider some physical attributes of the reactions (such as the barrier heights) in addition to the number of reactions while searching for pathways.

Most of previous work~\cite{graph_network3} related to searching for pathways in reaction networks represent the underlying network as a graph with simple edges. However, this does not capture the nature of chemical reactions faithfully as it might involve multiple reactant molecules and produce multiple product molecules. Often the modeling tracks the most important molecules in the reactions~\cite{yaks,yarp}, relegating the other molecules involved in the reaction as reagents or side-products (in the style of synthesis maps of natural products). However, since the mathematical framework of hypergraphs exists to represent many-to-many mappings, we choose to use it instead of graphs. This has only one drawback: the visualization of the reaction networks becomes a little more involved and non-intuitive.

Studies focusing on large reaction networks such as~\cite{shortest_path_hypergraph1,shortest_path_hypergraph2} often extract the network from databases or existing knowledge. In contrast, using a generative methodology which can create the network on the fly (from a set of input molecules and a set of rules representing classes of reactions) allows us to investigate regions of the molecular space what have not been studied before. Moreover, it removes biases from previous experiments and studies that might have crept in while curating the database. If a particular promising molecule is not entered into the database, then no pathways queried from that network will contain it.

While there exists prior work which includes the generation of reaction networks, these networks were mostly generated using the gradient of potential information~\cite{yarp,empi1}. The determination of the gradient information used in potential energy surface explorations is computationally expensive and might limit the size of the studied network. Such studies focus on expanding the network only in the most promising directions and do not consider the remaining molecular space. Such a choice necessitated by computational costs hinders a thorough examination of the molecular space for pathways.  Chemical heuristics used by~\cite{yaks} offers a way out of this problem, and our method uses reaction templates to generate the networks this other studies in this class. We aim to generate an exhaustive network that is an over-approximation of what actually occurs in the system. The use of reaction templates allows us to cover all reaction possibilities in the generated network with a manageable computational cost. The current method uses graph rewriting and double pushouts as the formal framework behind the expansion of the network, which to the best of our knowledge is not used in other similar works.

\section{Conclusion} \label{sec: conclusion}

In this submission, we have described a methodology for kinetically informed exploration of pathways in reaction networks. 
One part of the methodology is an automated workflow for annotating individual reactions with their energy barriers, computed using a diverse selection of tools---RDKit~\cite{rdkit_embedMolecule}, xTB, ASE~\cite{ase-paper} and NeuralNEB~\cite{neuralneb}. The goal of this workflow is to minimize the laborious manual interventions and intensive expert knowledge often necessary when executing quantum chemical calculations to determine energy barriers. 
As demonstrated in Section~\ref{sec: results}, the workflow can be applied on networks of rather large scale, can work on the fly with generative network construction, and uses relatively moderate computational resources.
The method does not take into account the interactions of solvent molecules while calculating the energy barriers. This and the question of how well models based on machine learning generalize past their training sets may lead to reasonable reservations on the precision of the estimates of the energy barriers for some use cases. However, our methodology offers the flexibility that this component can be substituted with alternatives, if desired.

As the other part of the proposed methodology, we formulated an ILP to query pathways in the annotated network and designed an objective function
for pathway evaluation. The query formalism is at the same time both precise and very flexible in what queries can be formulated. The objective function is made under the simplifying assumption that all molecules in the reaction network are present with unit concentrations at all times. Additionally, the costs assigned to pathways depend on the underlying reaction network considered, so comparisons across networks are not directly possible. However, the question of how to evaluate pathways is complex and the reader of course has the option to substitute a different objective function, based on other cost metrics. We do note, though, that our simplification and the formulation via ILP allows us to analyze much larger networks and using much less computational resources than alternatives like stochastic simulations or potential energy surface exploration.
We also note that by a progressive constraining of the ILP, we in the end can return a \emph{list} of high-scoring solutions, each representing structurally different pathways. Giving a list of good, structurally different solutions, rather than a single solution, helps mitigate disagreements about the details of pathway cost metrics.

Finally, we showcased the efficiency and utility of the described method on a concrete reaction network, studying pathways to pre-biotically significant molecules, such as glycine and 2-hydroxyethanoic acid from glycolonitrile, ammonia and water. In summary, our work contributes to field of computationally assisted analysis, design and elicidation of pathways in reaction networks represented as directed hypergraphs.

\section*{Acknowledgements}

This work was supported by generous funding from the European Union's Horizon 2021 Research and Innovation program under Marie Sklodowska-Curie Grant Agreement No.~101072930 (TACsy --- Training Alliance for Computational Systems Chemistry).

\begin{scriptsize}
    \bibliography{example_paper}

\begin{thebibliography}{40}
\providecommand{\natexlab}[1]{#1}
\providecommand{\url}[1]{\texttt{#1}}
\expandafter\ifx\csname urlstyle\endcsname\relax
  \providecommand{\doi}[1]{doi: #1}\else
  \providecommand{\doi}{doi: \begingroup \urlstyle{rm}\Url}\fi

\bibitem[Andersen(2024)]{mod_web}
Andersen, J.~L.
\newblock {M\O D}.
\newblock \url{http://mod.imada.sdu.dk}, 2024.

\bibitem[Andersen et~al.(2019)Andersen, Flamm, Merkle, and
  Stadler]{jakob_integer_hyperflows}
Andersen, J.~L., Flamm, C., Merkle, D., and Stadler, P.~F.
\newblock Chemical transformation motifs—modelling pathways as integer
  hyperflows.
\newblock \emph{IEEE/ACM Transactions on Computational Biology and
  Bioinformatics}, 16\penalty0 (2):\penalty0 510--523, 2019.
\newblock \doi{10.1109/TCBB.2017.2781724}.
\newblock URL \url{https://ieeexplore.ieee.org/document/8171738}.

\bibitem[Chang et~al.(2025)Chang, Tsai, and Li]{ml_reaction_barrier7}
Chang, H.-C., Tsai, M.-H., and Li, Y.-P.
\newblock Enhancing activation energy predictions under data constraints using
  graph neural networks.
\newblock \emph{Journal of Chemical Information and Modeling}, 65\penalty0
  (3):\penalty0 1367--1377, 2025.
\newblock \doi{10.1021/acs.jcim.4c02319}.
\newblock URL \url{https://doi.org/10.1021/acs.jcim.4c02319}.
\newblock PMID: 39862160.

\bibitem[Danger et~al.(2012)Danger, Duvernay, Theulé, Borget, and
  Chiavassa]{MeCN_precursor}
Danger, G., Duvernay, F., Theulé, P., Borget, F., and Chiavassa, T.
\newblock Hydroxyacetonitrile formation in astrophysical conditions.
  competition with the aminomethanol, a glycine precursor.
\newblock \emph{The Astrophysical Journal}, 756\penalty0 (1):\penalty0 11, aug
  2012.
\newblock \doi{10.1088/0004-637X/756/1/11}.
\newblock URL \url{https://dx.doi.org/10.1088/0004-637X/756/1/11}.

\bibitem[documentation()]{fba_alternative_pathways}
documentation, C.~T.
\newblock Flux balance analysis: Alternate optimal solutions.
\newblock
  \url{https://opencobra.github.io/cobratoolbox/stable/tutorials/tutorial_practical_alternateOptimalSolutions.html}.

\bibitem[Duan et~al.(2023)Duan, Du, Jia, and Kulik]{OA-ReactDiff}
Duan, C., Du, Y., Jia, H., and Kulik, H.~J.
\newblock Accurate transition state generation with an object-aware equivariant
  elementary reaction diffusion model.
\newblock \emph{Nature Computational Science}, 3\penalty0 (12):\penalty0
  1045--1055, Dec 2023.
\newblock \doi{10.1038/s43588-023-00563-7}.
\newblock URL \url{https://doi.org/10.1038/s43588-023-00563-7}.

\bibitem[Ebrahim et~al.(2013)Ebrahim, Lerman, Palsson, et~al.]{cobrapy}
Ebrahim, A., Lerman, J., Palsson, B., et~al.
\newblock {COBRA}py: {CO}nstraints-{B}ased {R}econstruction and {A}nalysis for
  {P}ython.
\newblock \emph{Syst Biol}, 7\penalty0 (74), August 2013.
\newblock \doi{10.1186/1752-0509-7-74}.
\newblock URL \url{https://doi.org/10.1186/1752-0509-7-74}.

\bibitem[Grzybowski et~al.(2023)Grzybowski, Badowski, Molga, and
  Szymkuć]{beam_search_pq}
Grzybowski, B.~A., Badowski, T., Molga, K., and Szymkuć, S.
\newblock Network search algorithms and scoring functions for advanced-level
  computerized synthesis planning.
\newblock \emph{WIREs Computational Molecular Science}, 13\penalty0
  (1):\penalty0 e1630, 2023.
\newblock \doi{10.1002/wcms.1630}.
\newblock URL \url{https://doi.org/10.1002/wcms.1630}.

\bibitem[Heald \& Kroll(2020)Heald and Kroll]{reported_network2}
Heald, C.~L. and Kroll, J.~H.
\newblock The fuel of atmospheric chemistry: Toward a complete description of
  reactive organic carbon.
\newblock \emph{Science Advances}, 6\penalty0 (6):\penalty0 eaay8967, 2020.
\newblock \doi{10.1126/sciadv.aay8967}.
\newblock URL \url{https://www.science.org/doi/abs/10.1126/sciadv.aay8967}.

\bibitem[Heinen et~al.(2021)Heinen, von Rudorff, and von
  Lilienfeld]{ml_energy_barrier3}
Heinen, S., von Rudorff, G.~F., and von Lilienfeld, O.~A.
\newblock Toward the design of chemical reactions: Machine learning barriers of
  competing mechanisms in reactant space.
\newblock \emph{The Journal of Chemical Physics}, 155\penalty0 (6):\penalty0
  064105, 08 2021.
\newblock ISSN 0021-9606.
\newblock \doi{10.1063/5.0059742}.
\newblock URL \url{https://doi.org/10.1063/5.0059742}.

\bibitem[Henkelman \& Jónsson(2000)Henkelman and Jónsson]{neb}
Henkelman, G. and Jónsson, H.
\newblock Improved tangent estimate in the nudged elastic band method for
  finding minimum energy paths and saddle points.
\newblock \emph{The Journal of Chemical Physics}, 113\penalty0 (22):\penalty0
  9978--9985, 12 2000.
\newblock ISSN 0021-9606.
\newblock \doi{10.1063/1.1323224}.
\newblock URL \url{https://doi.org/10.1063/1.1323224}.

\bibitem[Huet et~al.(2024)Huet, Devergne, Magrino, and Saitta]{glycine_pathway}
Huet, L., Devergne, T., Magrino, T., and Saitta, A.~M.
\newblock A new route to the prebiotic synthesis of glycine via ab initio-based
  machine learning calculations.
\newblock \emph{The Journal of Physical Chemistry Letters}, 15\penalty0
  (34):\penalty0 8697--8705, 2024.
\newblock \doi{10.1021/acs.jpclett.4c01954}.
\newblock URL \url{https://doi.org/10.1021/acs.jpclett.4c01954}.
\newblock PMID: 39159425.

\bibitem[Ida et~al.(2023)Ida, Kojima, and Hori]{graph_network3}
Ida, T., Kojima, H., and Hori, Y.
\newblock Predicting and analyzing organic reaction pathways by combining
  machine learning and reaction network approaches.
\newblock \emph{Chem. Commun.}, 59:\penalty0 12439--12442, 2023.
\newblock \doi{10.1039/D3CC03890D}.
\newblock URL \url{http://dx.doi.org/10.1039/D3CC03890D}.

\bibitem[{IUPAC}(2025{\natexlab{a}})]{arrhenius1}
{IUPAC}.
\newblock International union of pure and applied chemistry: Arrhenius
  equation, 2025{\natexlab{a}}.
\newblock URL \url{https://doi.org/10.1351/goldbook.A00446}.

\bibitem[{IUPAC}(2025{\natexlab{b}})]{arrhenius2}
{IUPAC}.
\newblock International union of pure and applied chemistry: Modified arrhenius
  equation, 2025{\natexlab{b}}.
\newblock URL \url{https://doi.org/10.1351/goldbook.M03963}.

\bibitem[{IUPAC}(2025{\natexlab{c}})]{eyring_equation}
{IUPAC}.
\newblock International union of pure and applied chemistry: Transition state
  theory, 2025{\natexlab{c}}.
\newblock URL \url{https://doi.org/10.1351/goldbook.T06470}.

\bibitem[Jain et~al.(2013)Jain, Ong, Hautier, Chen, Richards, Dacek, Cholia,
  Gunter, Skinner, Ceder, and Persson]{database_energy_barriers}
Jain, A., Ong, S.~P., Hautier, G., Chen, W., Richards, W.~D., Dacek, S.,
  Cholia, S., Gunter, D., Skinner, D., Ceder, G., and Persson, K.~A.
\newblock Commentary: The materials project: A materials genome approach to
  accelerating materials innovation.
\newblock \emph{APL Materials}, 1\penalty0 (1):\penalty0 011002, 07 2013.
\newblock ISSN 2166-532X.
\newblock \doi{10.1063/1.4812323}.
\newblock URL \url{https://doi.org/10.1063/1.4812323}.

\bibitem[Kowalik et~al.(2012)Kowalik, Gothard, Drews, Gothard, Weckiewicz,
  Fuller, Grzybowski, and Bishop]{search_crn}
Kowalik, M., Gothard, C.~M., Drews, A.~M., Gothard, N.~A., Weckiewicz, A.,
  Fuller, P.~E., Grzybowski, B.~A., and Bishop, K. J.~M.
\newblock Parallel optimization of synthetic pathways within the network of
  organic chemistry.
\newblock \emph{Angewandte Chemie International Edition}, 51\penalty0
  (32):\penalty0 7928--7932, 2012.
\newblock \doi{10.1002/anie.201202209}.
\newblock URL \url{https://doi.org/10.1002/anie.201202209}.

\bibitem[Krieger \& Kececioglu(2023)Krieger and
  Kececioglu]{shortest_path_hypergraph1}
Krieger, S. and Kececioglu, J.
\newblock Shortest hyperpaths in directed hypergraphs for reaction pathway
  inference.
\newblock \emph{Journal of Computational Biology}, 30\penalty0 (11):\penalty0
  1198--1225, 2023.
\newblock \doi{10.1089/cmb.2023.0242}.
\newblock URL \url{https://doi.org/10.1089/cmb.2023.0242}.
\newblock PMID: 37906100.

\bibitem[Larsen et~al.(2017)Larsen, Mortensen, Blomqvist, Castelli,
  Christensen, Dułak, Friis, Groves, Hammer, Hargus, Hermes, Jennings, Jensen,
  Kermode, Kitchin, Kolsbjerg, Kubal, Kaasbjerg, Lysgaard, Maronsson, Maxson,
  Olsen, Pastewka, Peterson, Rostgaard, Schiøtz, Schütt, Strange, Thygesen,
  Vegge, Vilhelmsen, Walter, Zeng, and Jacobsen]{ase-paper}
Larsen, A.~H., Mortensen, J.~J., Blomqvist, J., Castelli, I.~E., Christensen,
  R., Dułak, M., Friis, J., Groves, M.~N., Hammer, B., Hargus, C., Hermes,
  E.~D., Jennings, P.~C., Jensen, P.~B., Kermode, J., Kitchin, J.~R.,
  Kolsbjerg, E.~L., Kubal, J., Kaasbjerg, K., Lysgaard, S., Maronsson, J.~B.,
  Maxson, T., Olsen, T., Pastewka, L., Peterson, A., Rostgaard, C., Schiøtz,
  J., Schütt, O., Strange, M., Thygesen, K.~S., Vegge, T., Vilhelmsen, L.,
  Walter, M., Zeng, Z., and Jacobsen, K.~W.
\newblock The atomic simulation environment - a python library for working with
  atoms.
\newblock \emph{Journal of Physics: Condensed Matter}, 29\penalty0
  (27):\penalty0 273002, 2017.
\newblock URL \url{https://doi.org/10.1088/1361-648X/aa680e}.

\bibitem[Lee et~al.(2020)Lee, Woo~Kim, and Youn~Kim]{mc_trees}
Lee, K., Woo~Kim, J., and Youn~Kim, W.
\newblock Efficient construction of a chemical reaction network guided by a
  monte carlo tree search.
\newblock \emph{ChemSystemsChem}, 2\penalty0 (5):\penalty0 e1900057, 2020.
\newblock \doi{10.1002/syst.201900057}.
\newblock URL \url{https://doi.org/10.1002/syst.201900057}.

\bibitem[Liu et~al.(2021)Liu, Grinberg~Dana, Johnson, Goldman, Jocher, Payne,
  Grambow, Han, Yee, Mazeau, Blondal, West, Goldsmith, and Green]{rmg2}
Liu, M., Grinberg~Dana, A., Johnson, M.~S., Goldman, M.~J., Jocher, A., Payne,
  A.~M., Grambow, C.~A., Han, K., Yee, N.~W., Mazeau, E.~J., Blondal, K., West,
  R.~H., Goldsmith, C.~F., and Green, W.~H.
\newblock Reaction mechanism generator v3.0: Advances in automatic mechanism
  generation.
\newblock \emph{Journal of Chemical Information and Modeling}, 61\penalty0
  (6):\penalty0 2686--2696, 2021.
\newblock \doi{10.1021/acs.jcim.0c01480}.
\newblock URL \url{https://doi.org/10.1021/acs.jcim.0c01480}.
\newblock PMID: 34048230.

\bibitem[McDermott et~al.(2021)McDermott, Dwaraknath, and Persson]{CRN_qm}
McDermott, M.~J., Dwaraknath, S.~S., and Persson, K.~A.
\newblock A graph-based network for predicting chemical reaction pathways in
  solid-state materials synthesis.
\newblock \emph{Nat Commun}, 12, April 2021.
\newblock \doi{10.1038/s41467-021-23339-x}.
\newblock URL \url{https://doi.org/10.1038/s41467-021-23339-x}.

\bibitem[Nakamura et~al.(2012)Nakamura, Hachiya, Saito,
  et~al.]{shortest_path_hypergraph2}
Nakamura, M., Hachiya, T., Saito, Y., et~al.
\newblock An efficient algorithm for de novo predictions of biochemical
  pathways between chemical compounds.
\newblock \emph{Bioinformatics}, 13 (Suppl 17)\penalty0 (S8), December 2012.
\newblock \doi{10.1186/1471-2105-13-S17-S8}.
\newblock URL \url{https://doi.org/10.1186/1471-2105-13-S17-S8}.

\bibitem[N\o{}jgaard et~al.(2021)N\o{}jgaard, Fontana, Hellmuth, and
  Merkle]{atom_tracking_mod}
N\o{}jgaard, N., Fontana, W., Hellmuth, M., and Merkle, D.
\newblock Cayley graphs of semigroups applied to atom tracking in chemistry.
\newblock \emph{Journal of Computational Biology}, 28\penalty0 (7):\penalty0
  701--715, 2021.
\newblock \doi{10.1089/cmb.2020.0548}.
\newblock URL \url{https://doi.org/10.1089/cmb.2020.0548}.
\newblock PMID: 34115945.

\bibitem[{Open Babel documentation}({\natexlab{a}})]{openBabel_make3d}
{Open Babel documentation}.
\newblock {make3D() function in Open Babel}.
\newblock
  \url{https://open-babel.readthedocs.io/en/latest/UseTheLibrary/Python_PybelAPI.html\#pybel.Molecule.make3D},
  {\natexlab{a}}.

\bibitem[{Open Babel
  documentation}({\natexlab{b}})]{openBabel_systematicRotorSearch}
{Open Babel documentation}.
\newblock {SystematicRotorSearch() function in Open Babel}.
\newblock
  \url{https://open-babel.readthedocs.io/en/latest/3DStructureGen/SingleConformer.html\#id3},
  {\natexlab{b}}.

\bibitem[Pal et~al.(2024)Pal, Fagerberg, Andersen, Flamm, Dittrich, and
  Merkle]{thermoflow}
Pal, A., Fagerberg, R., Andersen, J.~L., Flamm, C., Dittrich, P., and Merkle,
  D.
\newblock Finding thermodynamically favorable pathways in reaction networks
  using flows in hypergraphs and mixed integer linear programming, 2024.
\newblock URL \url{https://arxiv.org/abs/2411.15900}.

\bibitem[{RDKit documentation}()]{rdkit_embedMolecule}
{RDKit documentation}.
\newblock {embedMolecule() function in RDKit}.
\newblock
  \url{https://www.rdkit.org/docs/source/rdkit.Chem.rdDistGeom.html\#rdkit.Chem.rdDistGeom.EmbedMolecule}.

\bibitem[Schreiner et~al.(2022{\natexlab{a}})Schreiner, Bhowmik, Vegge, Busk,
  and Winther]{transition1x}
Schreiner, M., Bhowmik, A., Vegge, T., Busk, J., and Winther, O.
\newblock Transition1x - a dataset for building generalizable reactive machine
  learning potentials.
\newblock \emph{Sci Data}, 9\penalty0 (779), 2022{\natexlab{a}}.
\newblock \doi{10.1038/s41597-022-01870-w}.
\newblock URL \url{https://doi.org/10.1038/s41597-022-01870-w}.

\bibitem[Schreiner et~al.(2022{\natexlab{b}})Schreiner, Bhowmik, Vegge,
  Jørgensen, and Winther]{neuralneb}
Schreiner, M., Bhowmik, A., Vegge, T., Jørgensen, P.~B., and Winther, O.
\newblock Neuralneb—neural networks can find reaction paths fast.
\newblock \emph{Machine Learning: Science and Technology}, 3\penalty0
  (4):\penalty0 045022, dec 2022{\natexlab{b}}.
\newblock \doi{10.1088/2632-2153/aca23e}.
\newblock URL \url{https://dx.doi.org/10.1088/2632-2153/aca23e}.

\bibitem[Sch{\"u}tt et~al.(2021)Sch{\"u}tt, Unke, and Gastegger]{painn}
Sch{\"u}tt, K., Unke, O., and Gastegger, M.
\newblock Equivariant message passing for the prediction of tensorial
  properties and molecular spectra.
\newblock In Meila, M. and Zhang, T. (eds.), \emph{Proceedings of the 38th
  International Conference on Machine Learning}, volume 139 of
  \emph{Proceedings of Machine Learning Research}, pp.\  9377--9388, ., 18--24
  Jul 2021. PMLR.
\newblock URL \url{https://proceedings.mlr.press/v139/schutt21a.html}.

\bibitem[Thrush \& Kua(2018)Thrush and Kua]{implemented_CRN}
Thrush, K.~L. and Kua, J.
\newblock Reactions of glycolonitrile with ammonia and water: A free energy
  map.
\newblock \emph{The Journal of Physical Chemistry A}, 122\penalty0
  (33):\penalty0 6769--6779, 2018.
\newblock \doi{10.1021/acs.jpca.8b05900}.
\newblock URL \url{https://doi.org/10.1021/acs.jpca.8b05900}.
\newblock PMID: 30063827.

\bibitem[van~der Schaft et~al.(2013)van~der Schaft, Rao, and
  Jayawardhana]{crn_dynamics2}
van~der Schaft, A., Rao, S., and Jayawardhana, B.
\newblock On the mathematical structure of balanced chemical reaction networks
  governed by mass action kinetics.
\newblock \emph{SIAM Journal on Applied Mathematics}, 73\penalty0 (2):\penalty0
  953--973, 2013.
\newblock \doi{10.1137/11085431X}.
\newblock URL \url{https://doi.org/10.1137/11085431X}.

\bibitem[Vijay et~al.(2022)Vijay, Kastlunger, Chan, and
  Nørskov]{scaling_limitations}
Vijay, S., Kastlunger, G., Chan, K., and Nørskov, J.~K.
\newblock Limits to scaling relations between adsorption energies?
\newblock \emph{The Journal of Chemical Physics}, 156\penalty0 (23):\penalty0
  231102, 06 2022.
\newblock ISSN 0021-9606.
\newblock \doi{10.1063/5.0096625}.
\newblock URL \url{https://doi.org/10.1063/5.0096625}.

\bibitem[Wang et~al.(2014)Wang, Titov, McGibbon, et~al.]{crn_expand_qm3}
Wang, L., Titov, A., McGibbon, R., et~al.
\newblock Discovering chemistry with an ab initio nanoreactor.
\newblock \emph{Nature Chem}, \penalty0 (6):\penalty0 1044–1048, November
  2014.
\newblock \doi{10.1038/nchem.2099}.
\newblock URL \url{https://doi.org/10.1038/nchem.2099}.

\bibitem[Woulfe \& Savoie(2025)Woulfe and Savoie]{yaks}
Woulfe, M. and Savoie, B.~M.
\newblock Chemical reaction networks from scratch with reaction prediction and
  kinetics-guided exploration.
\newblock \emph{Journal of Chemical Theory and Computation}, 21\penalty0
  (3):\penalty0 1276--1291, 2025.
\newblock \doi{10.1021/acs.jctc.4c01401}.
\newblock URL \url{https://doi.org/10.1021/acs.jctc.4c01401}.
\newblock PMID: 39883589.

\bibitem[{xTB documentation}()]{xtb_geomOpt}
{xTB documentation}.
\newblock {geometry optimization in xTB}.
\newblock \url{https://xtb-docs.readthedocs.io/en/latest/optimization.html}.

\bibitem[Zhang et~al.(2023)Zhang, Xu, and Lan]{empi1}
Zhang, Y., Xu, C., and Lan, Z.
\newblock Automated exploration of reaction networks and mechanisms based on
  metadynamics nanoreactor simulations.
\newblock \emph{Journal of Chemical Theory and Computation}, 19\penalty0
  (23):\penalty0 8718--8731, 2023.
\newblock \doi{10.1021/acs.jctc.3c00752}.
\newblock URL \url{https://doi.org/10.1021/acs.jctc.3c00752}.
\newblock PMID: 38031422.

\bibitem[Zhao \& Savoie(2021)Zhao and Savoie]{yarp}
Zhao, Q. and Savoie, B.
\newblock Simultaneously improving reaction coverage and computational cost in
  automated reaction prediction tasks.
\newblock \emph{Nat Comput Sci}, \penalty0 (1):\penalty0 479–490, July 2021.
\newblock \doi{10.1038/s43588-021-00101-3}.
\newblock URL \url{https://doi.org/10.1038/s43588-021-00101-3}.

\end{thebibliography}
    \bibliographystyle{icml2025}
\end{scriptsize}


\appendix
\label{appendix}

\section{On the Formalism for Representing Reaction networks}\label{first appendix}

While one might choose to represent a reaction formally by \eqref{chemical_equation}, for more practical scenarios, it is just a collection of reactant molecules (with their multiplicities) that gets converted into a collection of product molecules (with some different multiplicities). For example,

\begin{center}
\begin{tiny}
\begin{tabular}{ccccccc}
     $e_1$ &: &\chemfig{HO-[:30]-[:-30]~[:-30]N} &+ &\chemfig{H_2O} &$\longrightarrow$ &\chemfig{HO-[:-30]-[:30](-[:90]OH)=[:-30]NH} \\
     && $v_0$ && $v_2$ && $v_3$
\end{tabular}
\end{tiny}
\end{center}

This reaction is a two-to-one map which converts molecules $\{v_0, v_1\}$ to $\{v_3\}$. We choose to represent this map graphically as\\
\begin{center}
    \begin{tikzpicture}[transform canvas={scale=0.8}]   
        \tikzset{
            vertex/.style = {shape=circle, draw, minimum size = 2.0em},
            vertex2/.style = {shape=rectangle, draw, minimum size = 0.5em},    
            edge/.style = {->, -{Stealth[length = 5pt]}}
        }        
        \node[vertex] (v0) at  (0,1) {$v_0$};
        \node[vertex] (v2) at  (0,0) {$v_2$};
        \node[vertex2] (e1) at  (1.2,0.5) {\tiny{$e_1$}};
        \node[vertex] (v3) at  (2.5,0.5) {$v_3$};
        \draw[edge] (v0) to (e1);
        \draw[edge] (v2) to (e1);
        \draw[edge] (e1) to (v3);
    \end{tikzpicture}
\end{center}
\vspace{0.2cm}

This object is called an hyperedge and the individual molecules $v_0$, $v_2$ and $v_3$ form the vertices. However, in a practical scenario, when the molecules $v_0$ and $v_2$ are allowed to react, reaction $e_1$ is not the only reaction that would occur and the system would not stop reacting after producing $v_3$. Successive reactions might also be possible, some of which might be:

\begin{center}
\begin{tiny}
\begin{tabular}{cccccc}
     $e_1$: &\chemfig{H_2O} + &\chemfig{HO-[:30]-[:-30]~[:-30]N} &$\rightarrow$ &\chemfig{HO-[:-30]-[:30](-[:90]OH)=[:-30]NH} \\
     & $v_2$ & $v_0$ && $v_3$ \\
     $e_2$: &\chemfig{H_2O} + &\chemfig{HO-[:-30]-[:30](-[:90]OH)=[:-30]NH} &$\rightleftarrows$ &\chemfig{HO-[:-30]-[:30](-[:120]HO)(-[:60]OH)-[:-30]NH_2} \\
     & $v_2$ & $v_3$ && $v_4$ \\
     $e_3$: &&\chemfig{HO-[:-30]-[:30](-[:90]OH)=[:-30]NH} &$\rightleftarrows$ &\chemfig{HO-[:-30]-[:30](=[:90]O)-[:-30]NH_2}\\
     && $v_3$ && $v_5$ \\
     $e_4$: &\chemfig{H_2O} + &\chemfig{HO-[:-30]-[:30](=[:90]O)-[:-30]NH_2} &$\rightleftarrows$ &\chemfig{HO-[:-30]-[:30](-[:120]HO)(-[:60]OH)-[:-30]NH_2} \\
     & $v_2$ & $v_5$ && $v_4$ \\
     $e_5$: &&\chemfig{HO-[:-30]-[:30](-[:120]HO)(-[:60]OH)-[:-30]NH_2} &$\rightarrow$ &\chemfig{HO-[:-30]-[:30](=[:90]O)-[:-30]OH} &+ \chemfig{NH_3}\\
     && $v_4$  && $v_6$ & $v_1$ \\
     $e_6$: &&\chemfig{HO-[:-30]-[:30](-[:90]OH)=[:-30]NH} &$\rightleftarrows$ &\chemfig{HO-[:-30]=[:30](-[:90]OH)-[:-30]NH_2} \\
     && $v_3$ && $v_7$ \\
     $e_7$: &&\chemfig{HO-[:-30]-[:30](=[:90]O)-[:-30]NH_2} &$\rightleftarrows$ &\chemfig{HO-[:-30]=[:30](-[:90]OH)-[:-30]NH_2} \\
     && $v_5$ && $v_7$ \\
     $e_8$: &&\chemfig{HO-[:-30]=[:30](-[:90]OH)-[:-30]NH_2} &$\rightleftarrows$ &\chemfig{O=[:-30]-[:30](-[:90]OH)-[:-30]NH_2} \\
     && $v_7$ && $v_8$ \\
     $e_9$: &&\chemfig{O=[:-30]-[:30](-[:90]OH)-[:-30]NH_2} &$\rightleftarrows$ &\chemfig{O=[:-30]-[:30]=[:-30]O} &+\chemfig{NH_3} \\
     && $v_8$ && $v_9$ & $v_1$ \\
     $e_{10}$: &&\chemfig{O=[:-30]-[:30](-[:90]OH)-[:-30]NH_2} &$\rightleftarrows$ &\chemfig{O=[:-30]-[:30]=[:-30]NH} &+\chemfig{H_2O} \\
     && $v_8$ && $v_{10}$ & $v_2$ \\
     $e_{11}$: &\chemfig{H_2O}+ &\chemfig{O=[:-30]-[:30](-[:90]OH)-[:-30]NH_2} &$\rightleftarrows$ &\chemfig{HO-[:-30](-[:-90]OH)-[:30](-[:90]OH)-[:-30]NH_2} \\
     &$v_2$ & $v_8$ && $v_{11}$      
\end{tabular}
\end{tiny}
\end{center}

\begin{figure}[t]
    \centering
    \includegraphics[scale=0.8]{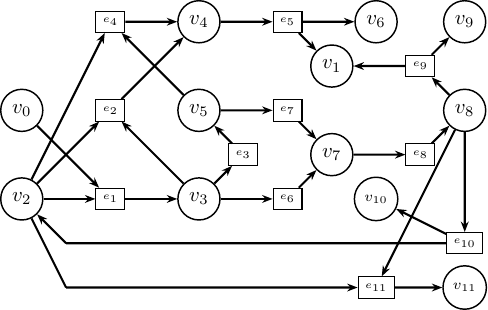}
    \caption{A hypergraph representation of the reactions in the concurrent system.}
    \label{fig:sample_network}
\end{figure}

Some of the reactions that might occur when molecules of $v_0$ and $v_2$ are allowed to react is shown. To a chemist, not all the reactions would look feasible due to reactivity constraints on the molecules, which would disallow some of the reactions. However, all these reactions might possibly occur in the system in parallel, making it an example of a concurrent system. We would represent each reaction by a separate hyperedge and each class of molecules by a distinct vertex. This results in the structure shown in Figure~\ref{fig:sample_network}. The reverse reactions are omitted from the hypergraph in Figure~\ref{fig:sample_network} to reduce clutter. However, in our actual modeling, we do add separate hyperedges to represent the forward and backward directions of reversible reactions.

This data structure representing the concurrent set of reactions in the system is called a multi-directed hypergraph or for simplicity, just a hypergraph. An advantage of using this data structure is that is faithfully represents the many-to-many mappings which are present in reactions which are not unimolecular. Contrast this with a simple directed graph containing edges which map one vertex to another vertex, which are often used to represent and study diverse networks in many situations. The hypergraph in Figure~\ref{fig:hypergraph} is an abstraction of another system of concurrent reactions.

One observes that glyoxal is represented as vertex $v_9$ in the hypergraph. One might query, how one inside this hypergraph can find a pathway producing glyoxal while using the input molecules $v_0$ and $v_2$. One possible transformation pathway would be:
\begin{center}
	\begin{tiny}
		\begin{tabular}{ccc}
			 \chemfig{HO-[:30]-[:-30]~[:-30]N} $\xrightarrow{+\chemfig{H_2O}}$ &\chemfig{HO-[:-30]-[:30](-[:90]OH)=[:-30]NH} $\xrightarrow{\text{taut.}}$ &\chemfig{HO-[:-30]=[:30](-[:90]OH)-[:-30]NH_2}\\
			 $\xrightarrow{\text{taut.}}$ \chemfig{O=[:-30]-[:30](-[:90]OH)-[:-30]NH_2} & $\xrightarrow{-\chemfig{NH_3}}$ \chemfig{O=[:-30]-[:30]=[:-30]O}
		\end{tabular}
	\end{tiny}
\end{center}
\begin{figure}[h]
    \centering
    \includegraphics[scale=0.8]{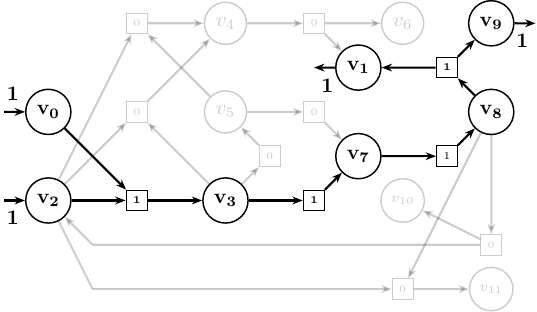}
    \caption{A pathway in a hypergraph with the queried molecules $v_9$. The number in the squares denote the number of times the reaction is used to carry out the transformation or the flow assigned to the corresponding hyperedge.}
    \label{fig:sample_path1}
\end{figure}

Highlighting the hyperedges used in this transformation pathway in the hypergraph results in Figure~\ref{fig:sample_path1}. This sequence of reactions in a pathway can be found by formulating it as a flow query with molecules $v_0$ and $v_2$ as the source molecules and $v_9$ as the target molecule. The ILP assigns integers to the hyperedges which denote the number of times the hyperedge is used in the pathway. Since the flows assigned to the hyperedges are constrained to be integers, this is called an integer hyperflow. The motivation behind using allowing only non-negative integers as permissible hyperflows on a hyperedge is discussed in Section~\ref{sec: discussions}.

Often, there might exist alternative solutions to the flow query. In this case, there exists another valid integer hyperflow to the same flow query in the network with molecules $v_0$ and $v_2$ as the source molecules and $v_9$ as the target molecule.
\begin{figure}[t]
    \centering
    \includegraphics[scale=0.8]{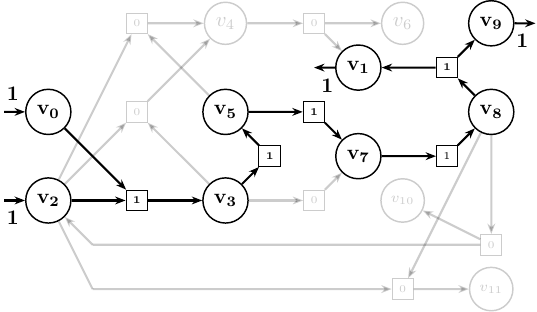}
    \caption{An alternative pathway to the same query as in Figure~\ref{fig:sample_path1}.}
    \label{fig:ssample_path2}
\end{figure}
It might so happen than the reaction $v_3 \rightarrow v_7$ is slow (because it has a high energy barrier) and the two step process $v_3 \rightarrow v_5 \rightarrow v_7$ is faster. To judge which of the two pathways is better, a score is assign to them by evaluating an objective function (expression \eqref{proposed_objective} is used in this work) on the pathway, and the one with a better score is ranked higher. This automated search for pathways in a reaction system represented as a hypergraph is the motivation behind introducing the formalism of integer hyperflows in Section~\ref{sec:hyperflows}.

\section{Generation of Reaction networks}\label{second appendix}

There exists a diverse range of reactions. However, they are often grouped together on the basis of their mechanisms to create classes of reactions. Reactions in a class share some properties: the reactive core (type of atoms taking part in the reaction), the bonds that are broken, the movement of elections, and the new bonds that are formed. We choose to represent these classes of reactions using reaction templates. Each template defines a change in the bondset of the molecules that take part in the reaction.

For example, in the list of reaction from Appendix~\ref{first appendix}:
\begin{itemize}
    \item reactions $e_3$, $e_5$, $e_6$ $e_7$ and $e_8$ have a 1,3-shift of a hydrogen atom (often called tautomerism).
    \item reactions $e_2$, $e_4$ and $e_{11}$ are additions to an trigonal planar carbon to produce a tetrahedral carbon.
    \item reactions $e_9$ and $e_{10}$ eliminate a small molecule from a carbon with two or more adjacent electronegative atoms to form a trigonal planar carbon.
    \item reaction $e_1$ is an addition to a linear carbon to produce a trigonal carbon.
\end{itemize}
Some might prefer more fine-grained reaction templates. For instance, one might define separate templates for keto-enol tautomerism, amide-iminol tautomerism or imine-enamine tautomerism, and similarly for addition to a trigonal planar carbonyl carbon atom and addition to a trigonal planar imine carbon atom. However, that is simply a question of the level of abstraction that the user chooses to work with.

\subsection{Graph transformation rules}
We choose to represent molecules using undirected labeled graphs (hereafter called molecular graphs). The nodes in the graph represent atoms labeled by the symbol of the element, and can carry attributes such as the charge on the atom. The edges in the graph represent bonds labeled by the bond type (single, double, triple or aromatic). with this representation of molecules, the reaction templates used to abstract reactions become graph transformation rules.

A graph transformation rule consists of a subgraph on the left side which can be replaced by the subgraph on the right side when a match of the left-hand side is found in an already existing molecular graph in the network. In other words, the rule converts a part of the graph (representing a fragment of a molecule) and performs a chemical reaction. Let's consider the exchange of hydrogen atoms between oxygen atoms in a carboxylic acid.
\begin{center}
\begin{tiny}
\begin{tabular}{ccc}
     \chemfig{HO-[:30]([:90]=O)-[:-30]R} &$\rightarrow$ &\chemfig{O=[:30]([:90]-OH)-[:-30]R} 
\end{tabular}
\end{tiny}
\end{center}

The graph transformation rule for this class of reactions would be:
\begin{center}
    \begin{tikzpicture}[scale=1.0]    
        \tikzset{
            arrow/.style = {->, -{Stealth[length = 5pt]}},
            bond/.style={line width=0.6pt},
            dbond/.style={double, double distance=2pt}
        }
        \node (c0) at (0,0) {C};
        \node (o1) at (1,0) {\textcolor{red}{O}};
        \node (o2) at (0,-1) {\textcolor{red}{O}};
        \node (h3) at (1,-1) {H};
        \draw[bond] (c0) to (o2);
        \draw[bond] (o2) to (h3);
        \draw[dbond] (c0) to (o1);

        \draw[arrow] (2,-0.5) to (3,-0.5);

        \node (c0) at (4,0) {C};
        \node (o1) at (5,0) {\textcolor{red}{O}};
        \node (o2) at (4,-1) {\textcolor{red}{O}};
        \node (h3) at (5,-1) {H};
        \draw[dbond] (c0) to (o2);
        \draw[bond] (o1) to (h3);
        \draw[bond] (c0) to (o1);
    \end{tikzpicture}
\end{center}

The amide-iminol conversion is a more substantial example.
\begin{center}
\begin{tiny}
\begin{tabular}{ccc}
     \chemfig{H_2N-[:30]([:90]=O)-[:-30]R} &$\rightarrow$ &\chemfig{HN=[:30]([:90]-OH)-[:-30]R} 
\end{tabular}
\end{tiny}
\end{center}

The graph transformation rule for this class of reactions would be:
\begin{center}
    \begin{tikzpicture}[scale=1.0]    
        \tikzset{
            arrow/.style = {->, -{Stealth[length = 5pt]}},
            bond/.style={line width=0.6pt},
            dbond/.style={double, double distance=2pt}
        }
        \node (c0) at (0,0) {C};
        \node (o1) at (1,0) {\textcolor{red}{O}};
        \node (n2) at (0,-1) {\textcolor{blue}{N}};
        \node (h3) at (1,-1) {H};
        \draw[bond] (c0) to (n2);
        \draw[bond] (n2) to (h3);
        \draw[dbond] (c0) to (o1);

        \draw[arrow] (2,-0.5) to (3,-0.5);

        \node (c0) at (4,0) {C};
        \node (o1) at (5,0) {\textcolor{red}{O}};
        \node (n2) at (4,-1) {\textcolor{blue}{N}};
        \node (h3) at (5,-1) {H};
        \draw[dbond] (c0) to (n2);
        \draw[bond] (o1) to (h3);
        \draw[bond] (c0) to (o1);
    \end{tikzpicture}
\end{center}
This rule rewrites an amide fragment when it is found in a molecule with an iminol fragment, thus effecting the required reaction.

For a keto-enol tautomerism like
\begin{center}
\begin{tiny}
\begin{tabular}{ccc}
     \chemfig{R'-[:-30](-[:-90]H)-[:30]([:90]=O)-[:-30]R} &$\rightarrow$ &\chemfig{R'-[:-30]=[:30]([:90]-OH)-[:-30]R} 
\end{tabular}
\end{tiny}
\end{center}

the graph transformation rule for this class of reactions would be:
\begin{center}
    \begin{tikzpicture}[scale=1.0]    
        \tikzset{
            arrow/.style = {->, -{Stealth[length = 5pt]}},
            bond/.style={line width=0.6pt},
            dbond/.style={double, double distance=2pt}
        }
        \node (c0) at (0,0) {C$^1$};
        \node (o1) at (1,0) {\textcolor{red}{O}};
        \node (c2) at (0,-1) {C$^2$};
        \node (h3) at (1,-1) {H};
        \draw[bond] (c0) to (c2);
        \draw[bond] (c2) to (h3);
        \draw[dbond] (c0) to (o1);

        \draw[arrow] (2,-0.5) to (3,-0.5);

        \node (c0) at (4,0) {C$^1$};
        \node (o1) at (5,0) {\textcolor{red}{O}};
        \node (c2) at (4,-1) {C$^2$};
        \node (h3) at (5,-1) {H};
        \draw[dbond] (c0) to (c2);
        \draw[bond] (o1) to (h3);
        \draw[bond] (c0) to (o1);
    \end{tikzpicture}
\end{center}
Note that the graph transformation rule indicates the bare minimum that is required to be present as a subgraph in the molecule for the reaction to occur. For example, the carbon at the 2-position, might have two adjacent hydrogen atoms, but the rule specifies that at least one hydrogen atom is required for the keto isomer to convert into the enol isomer. Without any hydrogen at the 2-position, the fragment on the left is not a subgraph of the molecule and this transformation rule cannot be applied to the graph.

Also, note that the objective here is not to capture the exact mechanism of the reaction, but rather specify what the molecular fragment is before and after the reaction. In all the three reactions mentioned above, the 1,3-hydrogen transfer involves a solvent molecule. In most cases, it is not the exact same hydrogen atom that would be in the molecule before and after the reaction.

The similarity in the graph transformation rule for the three reactions mentioned above above look similar. For a more elegant formulation, we define the following graph transformation rule to account for all the three cases above:
\begin{center}
    \begin{tikzpicture}[scale=1.0]    
        \tikzset{
            arrow/.style = {->, -{Stealth[length = 5pt]}},
            bond/.style={line width=0.6pt},
            dbond/.style={double, double distance=2pt}
        }
        \node (c0) at (0,0) {C};
        \node (o1) at (1,0) {$\mathcal{Y}$};
        \node (c2) at (0,-1) {$\mathcal{X}$};
        \node (h3) at (1,-1) {H};
        \draw[bond] (c0) to (c2);
        \draw[bond] (c2) to (h3);
        \draw[dbond] (c0) to (o1);

        \draw[arrow] (2,-0.5) to (3,-0.5);

        \node (c0) at (4,0) {C};
        \node (o1) at (5,0) {$\mathcal{Y}$};
        \node (c2) at (4,-1) {$\mathcal{X}$};
        \node (h3) at (5,-1) {H};
        \draw[dbond] (c0) to (c2);
        \draw[bond] (o1) to (h3);
        \draw[bond] (c0) to (o1);

        \node at (2.5,-1.5) {$\mathcal{X}, \mathcal{Y} \in \{\text{C, N, O}\}$};
    \end{tikzpicture}
\end{center}
This graph transformation rule also encodes an allyl hydrogen shift, an imine-enamine tautomerism, or the hydrogen exchange in amidine.
\begin{center}
\begin{tiny}
\begin{tabular}{ccc}
    \chemfig{R_4-[:-30](-[:120]R_5)(-[:-90]H)-[:30]([:90]-R_3)=[:-30](-[:-90]R_2)-[:30]R_1} &$\rightarrow$ &\chemfig{R_4-[:-30](-[:-90]R_5)=[:30]([:90]-R_3)-[:-30](-[:-90]R_2)(-[:60]H)-[:30]R_1} \\
    \chemfig{R'-[:-30](-[:-90]H)-[:30]([:90]=NH)-[:-30]R} &$\rightarrow$ &\chemfig{R'-[:-30]=[:30]([:90]-NH_2)-[:-30]R} \\
    \chemfig{H_2N-[:30]([:90]=NH)-[:-30]R} &$\rightarrow$ &\chemfig{HN=[:30]([:90]-NH_2)-[:-30]R} 
\end{tabular}
\end{tiny}
\end{center}

However, all these six different reactions are expected to have differing mechanisms. This graph transformation rule introduces two variables $\mathcal{X}, \mathcal{Y}$ which can be matched with the label C, N or O, or in other words the atoms in the place of  $\mathcal{X}, \mathcal{Y}$ in the molecular fragment can be a carbon, nitrogen or oxygen atom.

Similarly the graph transformation rule
\begin{center}
    \begin{tikzpicture}[scale=1.0]    
        \tikzset{
            arrow/.style = {->, -{Stealth[length = 5pt]}},
            bond/.style={line width=0.6pt},
            dbond/.style={double, double distance=2pt}
        }
        \node (c0) at (0,-1) {C};
        \node (o1) at (0,0) {$\mathcal{Y}$};
        \node (c2) at (1,-1) {$\mathcal{X}$};
        \node (h3) at (1,0) {H};
        \draw[bond] (c2) to (h3);
        \draw[dbond] (c0) to (o1);

        \draw[arrow] (2,-0.5) to (3,-0.5);

        \node (c0) at (4,-1) {C};
        \node (o1) at (4,0) {$\mathcal{Y}$};
        \node (c2) at (5,-1) {$\mathcal{X}$};
        \node (h3) at (5,0) {H};
        \draw[dbond] (c0) to (c2);
        \draw[bond] (o1) to (h3);
        \draw[bond] (c0) to (o1);

        \node at (2.5,-1.5) {$\mathcal{X}, \mathcal{Y} \in \{\text{N, O}\}$};
    \end{tikzpicture}
\end{center}
encodes the addition of a water or ammonia molecule to a carbonyl or imine group.
\begin{center}
\begin{tiny}
\begin{tabular}{ccc}
    \chemfig{R'-[:30]([:90]=NH)-[:-30]R} &$\xrightarrow{+\chemfig{NH_3}}$ &\chemfig{R'-[:30]([:90]-NH_2)(-[:-60]NH_2)-[:-30]R} \\
    \chemfig{R'-[:30]([:90]=NH)-[:-30]R} &$\xrightarrow{+\chemfig{H_2O}}$ &\chemfig{R'-[:30]([:90]-NH_2)(-[:-60]OH)-[:-30]R} \\
    \chemfig{R'-[:30]([:90]=O)-[:-30]R} &$\xrightarrow{+\chemfig{NH_3}}$ &\chemfig{R'-[:30]([:90]-OH)(-[:-60]NH_2)-[:-30]R} \\
     \chemfig{R'-[:30]([:90]=O)-[:-30]R} &$\xrightarrow{+\chemfig{H_2O}}$ &\chemfig{R'-[:30]([:90]-OH)(-[:-60]OH)-[:-30]R} 
\end{tabular}
\end{tiny}
\end{center}
We admit that these reactions are not a simple one step process as the transformation rule make them appear to be and not all of these possible reactions is always feasible for all reactant molecules.

In all the graph transformation rules in this section, the context graph $K$, which comes between the left and right-hand sides, is omitted. The context graph must be added to the graph transformation rules for mathematical correctness. It ensures that logically the bonds are first removed, and then added, making the graph rewriting a two-step process.

\subsection{Rule application to a graph}

\begin{figure}[th]
    \centering
    \includegraphics[width=0.8\linewidth]{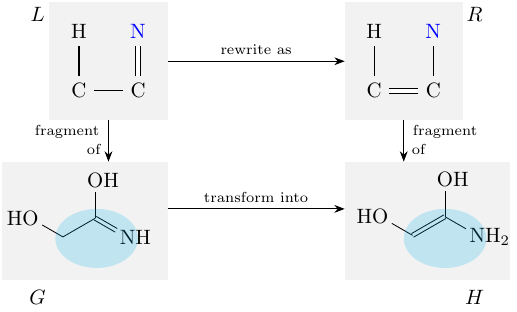}
    \begin{tikzpicture}
        \draw (0, 0) -- (4, 0);
    \end{tikzpicture}
    \vspace{1.0cm}\\
    \includegraphics[width=0.8\linewidth]{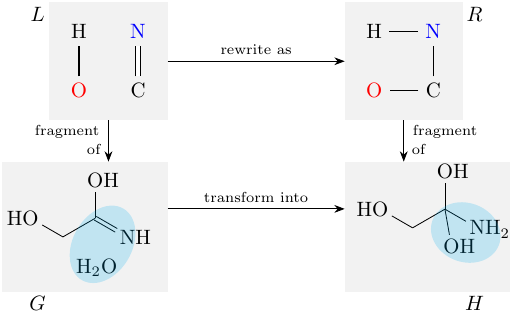}
    \caption{Two examples of the application of a graph transformation rule to a collection of molecules.}
    \label{fig:rule_application1}
\end{figure}

When a graph transformation rule is applied to a set of reactant molecules, a reaction might occur. First, it is checked if the left-hand side of the rule, $L$, is present as a subgraph in the set of reactant molecules $G$. If so, then that subgraph is rewritten according to the rule to R, to generate a new set of molecules $H$, the product molecules. This process is illustrated in Figure~\ref{fig:rule_application1}.

\begin{figure}
    \centering
    \includegraphics[width=0.8\linewidth]{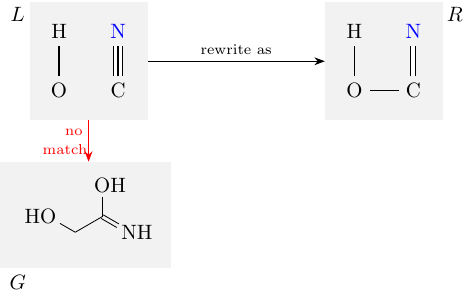}
    \caption{A case where the graph transformation rule cannot be applied to a molecule}
    \label{fig:rule_application2}
\end{figure}

If the left-hand side of the rule, $L$, is not present in the set of molecules $G$, then that reaction template cannot be applied to the given set of molecules as illustrated in Figure~\ref{fig:rule_application2}.

Therefore, the application of a rule to an existing set of molecules might produce new molecules, which in turn expands the molecular space. Repeated application of these rules constructs a network, which is detailed in the next section.

\subsection{Construction of the reaction network}

The construction of the reaction network requires a set of input molecules and some graph transformation rules. A match between a collection of molecules and the subgraph on the left-hand side of one of the transformation rules means that this reaction template can be applied to that collection of molecules created from the input set of molecules. Thus, when a rule is applied to a collection of molecules from the current set of molecules, new molecules are created. In a generative process, these new molecules are then added to the set of available molecules (i.e., this set is updated). Graph transformation rules can be recursively applied to this new set of molecules, effecting more reactions to be made and thus the molecular space expands. The hypergraph representation of the network keeps track of all the reactions performed and of the molecules current present in the network.

\begin{figure}
    \centering
    \includegraphics[scale=0.25]{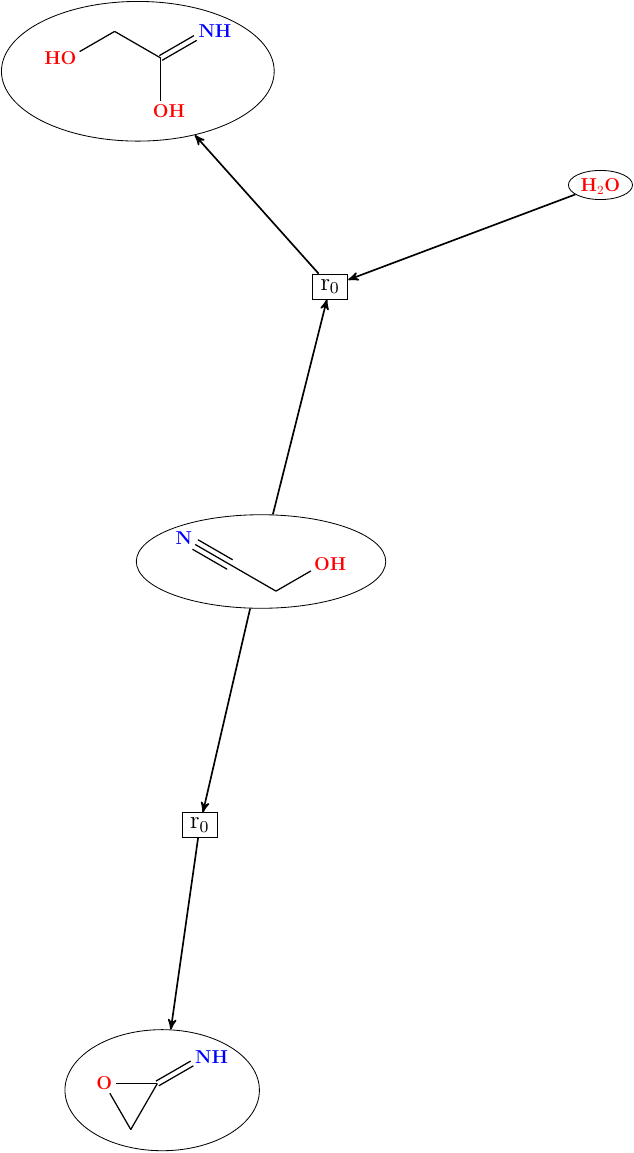}
    \hspace{0.2cm}
    \includegraphics[scale=0.25]{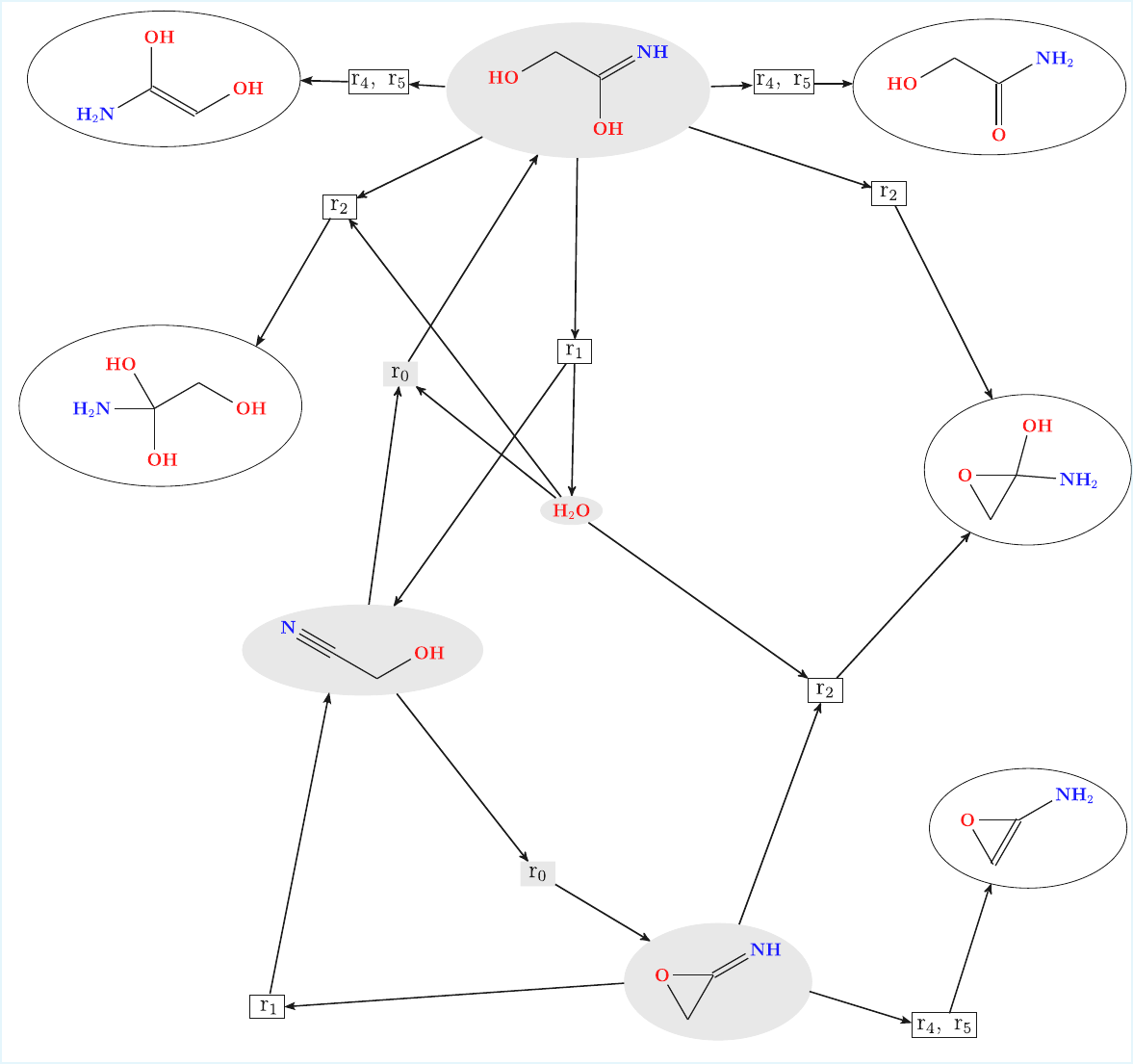}
    \caption{(left) The reaction network (with $4$ molecules and $2$ reactions) after one recursive application of the graph transformation rules, and (right) the reaction network (with $9$ molecules and $10$ reactions) after two recursive applications of the graph transformation rules. The part of the network already generated previously has been shaded.}
    \label{fig: grow_network}
\end{figure}

\begin{figure}[h]
    \centering
    \includegraphics[scale=0.25]{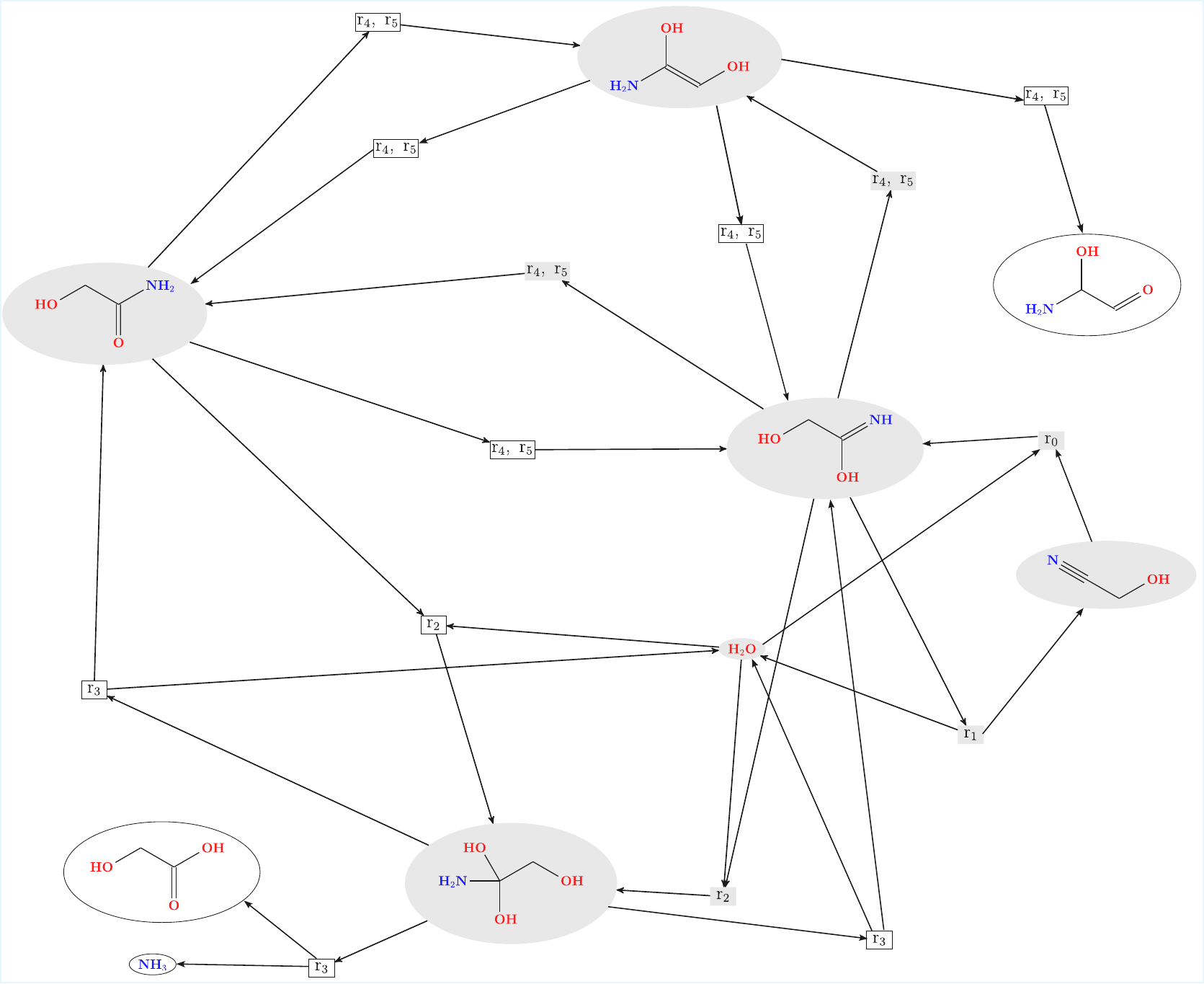}
    \caption{The reaction network (with $9$ molecules and $14$ reactions) after three recursive applications of the graph transformation rules with the 3-cycles filtered out. The part of the network already generated previously has been shaded.}
    \label{fig: recurse3}
\end{figure}

\begin{figure}[thb]
    \centering
    \includegraphics[scale=0.15]{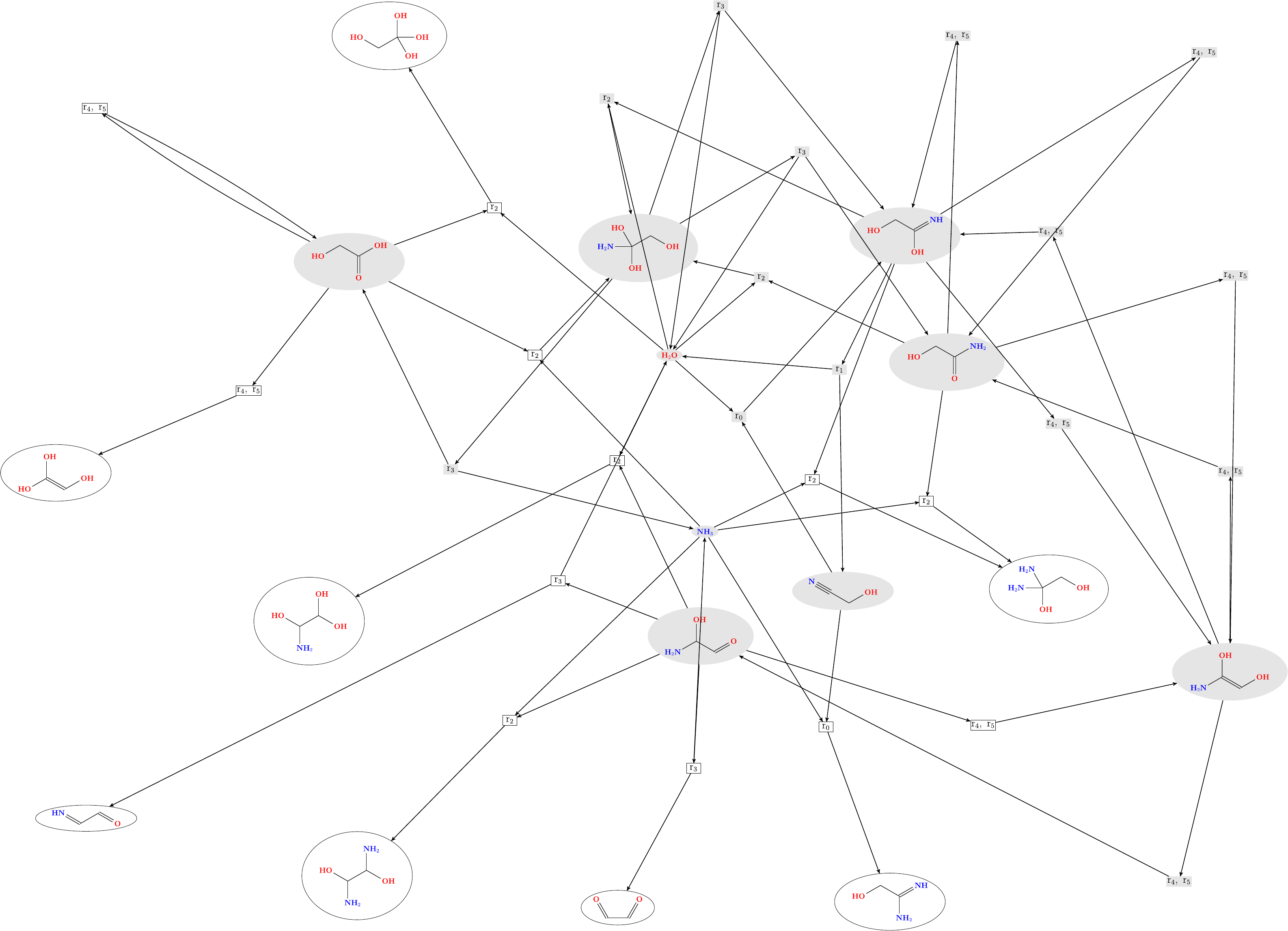}
    \caption{The reaction network (with $17$ molecules and $26$ reactions) after four recursive applications of the graph transformation rules with the 3-cycles filtered out. The part of the network that was already generated aftre three recersive steps has been shaded.}
    \label{fig: recurse4}
\end{figure}

The result of the recursive application of the graph transformation rules from Appendix~\ref{third appendix} to the input molecules water and 2-hydroxyethanenitrile is shown in Figures~\ref{fig: grow_network}, \ref{fig: recurse3} and \ref{fig: recurse4}.

Notice that some molecules in Figure~\ref{fig: grow_network}, such as the 3-cycles, are not realistic. One can disallow their addition to the generated reaction network by using so-called right predicates. Right predicates filter out certain reactions from the network based on the properties of the molecular graphs that appear on the right-hand side of a derivation. This, it prunes the generated network and makes its growth more manageable and potentially more realistic. The effect of using the right predicate can be seen in the hypergraph in Figure~\ref{fig: recurse3}, which has $9$ molecules after three recursive expansion steps, while the hypergraph in Figure~\ref{fig: grow_network} already had $9$ molecules after two recursive steps when the right predicate was not used.

Often, one direction of a reaction is added to the hypergraph (adding new molecules to the existing set of molecules) and the reverse direction is added in the next recursion. This adds a new reaction between reactants and products both of which are already present in the set of existing molecules.
The graph transformation rules can be applied for a specified number of iterations, or until a specified limit on the size of the different molecules generated is reached which terminated the network generation process. This makes the derived hypergraph finite.

\section{Graph Transformation Rules Used}\label{third appendix}

The graph transformation rules used to expand the reaction network are listed in this appendix. The subgraph $L$ denotes a fragment in the reactant  molecules and the subgraph $R$ denotes a fragment in the product molecules. The subgraph $K$ is included as context for the sake of mathematical correctness. Since these graph transformation rules represent a class of reactions, they can be considered as reaction templates.

\newcommand\ruleScale{0.8}

\begin{figure}[!h]
    \begin{subfigure}{0.5\textwidth}
        \centering
        \dpoRule[scale=\ruleScale]{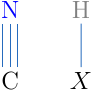}{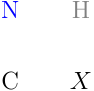}{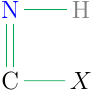}
        \caption{Rule 1: Addition of water or ammonia to nitrile group where $X \in \{\mathrm{N}, \mathrm{O}\}$ to form an imine. The reverse of this rule was used while expanding the hypergraph.}
        \label{fig:rule1}
    \end{subfigure}
    \begin{subfigure}{0.5\textwidth}
        \centering
        \dpoRule[scale=\ruleScale]{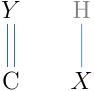}{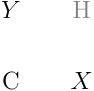}{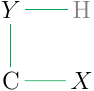}
	    \caption{Rule 2: Addition of water or ammonia to a double bond where $X, Y \in \{\mathrm{N}, \mathrm{O}\}$ (in an imine or carbonyl group respectively). The reverse of this rule was also used while expanding the hypergraph.}
        \label{fig:rule2}
    \end{subfigure}
    \begin{subfigure}{0.5\textwidth}
        \centering
        \dpoRule[scale=\ruleScale]{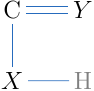}{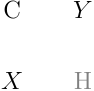}{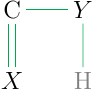}
	    \caption{Rule 3: 1,2 to 2,3 bond shift of the double bond to form tautomers where $X, Y \in \{\mathrm{N}, \mathrm{O}\}$. The reverse of this rule was also used while expanding the hypergraph.}
        \label{fig:rule3}
    \end{subfigure}
    \caption{Graph transformation rules (or reaction templates) used to generate the reaction network \ref{fig: hypergraph_for_RN} in which pathways were queried. The reverse of all the rules \ref{fig:rule1}, \ref{fig:rule2} and \ref{fig:rule3} were used while expanding the hypergraph.}
\end{figure}

\begin{figure}[!h]
    \centering
    \includegraphics[width=\linewidth]{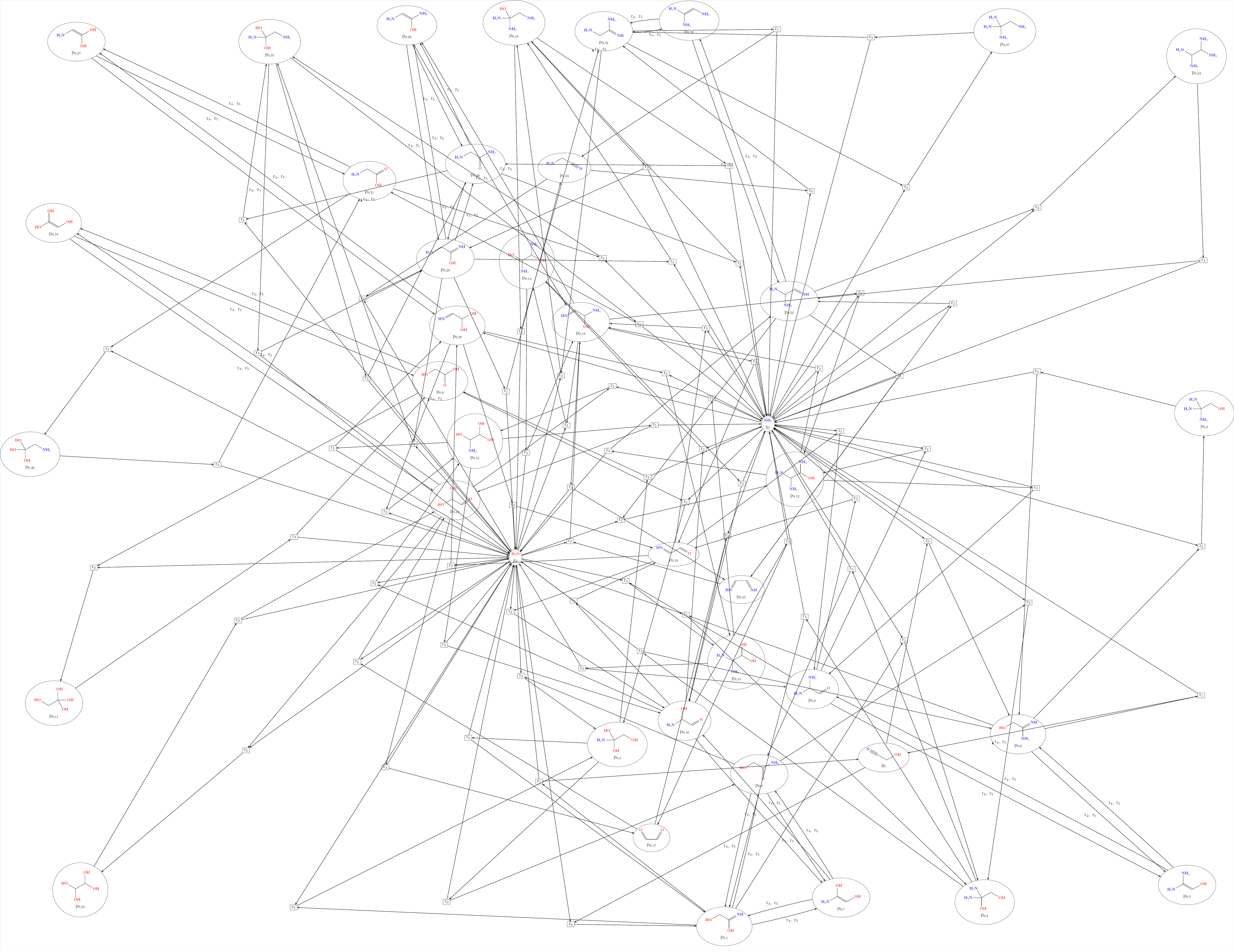}
    \caption{Hypergraph with $|V| = 67$ and $|E| = 202$, representing the reaction network considered in this work, provided here for the sake of completeness.}
    \label{fig: hypergraph_for_RN}
\end{figure}

The constructed reaction network is shown in Figure \ref{fig: hypergraph_for_RN}.

\section{List of Molecules in the Presented Pathways}

In this appendix, we supply the molecular structure and the corresponding keys of the vertices (it the form $v_i$, where $i$ is some integer) in the pathways that are presented in Section \ref{sec: results}. The two columns on the left side of Table \ref{tab:molecular_structure} depict the molecules in the pathways presented in Figure \ref{fig:pathway1}, while the column on the right side of Table \ref{tab:molecular_structure} depicts the molecules in the pathways presented in Figure \ref{fig:pathway2}.

\begin{table*}[t]
    \centering
    \begin{scriptsize}
    \begin{tabular}[t]{ll}
        \toprule
        Name & Molecular Structure \\
        \midrule
        $v_0$ & \chemfig{HO-[:30]-[:-30]~[:-30]N} \\ 
        $v_2$ & \chemfig{H_2O} \\
        $v_3$ & \chemfig{HO-[:30]-[:-30](-[:-90]OH)=[:30]NH} \\[0.4cm]
        $v_4$ & \chemfig{HO-[:30]=[:-30](-[:-90]OH)-[:30]NH_2} \\[0.4cm]
        $v_5$ & \chemfig{O=[:30]-[:-30](-[:-90]OH)-[:30]NH_2} \\[0.4cm]
        $v_6$ & \chemfig{O=[:30]-[:-30]=[:30]NH} \\
        $v_7$ & \chemfig{HO-[:30](-[:90]OH)-[:-30]=[:30]NH} \\
        $v_8$ & \chemfig{HO-[:30](-[:90]OH)=[:-30]-[:30]NH_2} \\
        $v_9$ & \chemfig{HO-[:30](=[:90]O)-[:-30]-[:30]NH_2} \\
        \bottomrule
    \end{tabular}
    \end{scriptsize}
    \hspace{1cm}
    \begin{scriptsize}
    \begin{tabular}[t]{ll}
        \toprule
        Name & Molecular Structure \\
        \midrule
        $v_1$ & \chemfig{NH_3} \\
        $v_{13}$ & \chemfig{HO-[:30]-[:-30](-[:-90]NH_2)=[:30]NH} \\[0.55cm]
        $v_{14}$ & \chemfig{HO-[:30]=[:-30](-[:-90]NH_2)-[:30]NH_2} \\[0.6cm]
        $v_{15}$ & \chemfig{O=[:30]-[:-30](-[:-90]NH_2)-[:30]NH_2} \\[0.5cm]
        $v_{16}$ & \chemfig{HO-[:30](-[:90]OH)-[:-30](-[:-90]NH_2)-[:30]NH_2} \\[0.5cm]
        $v_{36}$ & \chemfig{H_2N-[:30](-[:90]OH)-[:-30](-[:-90]OH)-[:30]NH_2} \\[0.4cm]
        $v_{52}$ & \chemfig{H_2N-[:30](=[:90]O)-[:-30]-[:30]OH} \\
        \bottomrule
    \end{tabular}
    \hspace{2cm}
    \begin{tabular}[t]{ll}
        \toprule
        Name & Molecular Structure \\
        \midrule
        $v_0$ & \chemfig{HO-[:30]-[:-30]~[:-30]N} \\ [0.3cm]
        $v_3$ & \chemfig{HO-[:30]-[:-30](-[:-90]OH)=[:30]NH} \\[0.4cm]
        $v_6$ & \chemfig{HO-[:30]-[:-30](-[:-60]OH)(-[:-120]HO)-[:30]NH_2} \\[0.5cm]
        $v_7$ & \chemfig{HO-[:30]-[:-30](=[:-90]O)-[:30]OH} \\
        $v_{15}$ & \chemfig{HO-[:30]-[:-30](=[:-90]O)-[:30]NH_2} \\
        $v_{24}$ & \chemfig{HO-[:30]=[:-30](-[:-90]OH)-[:30]NH_2} \\
        $v_{33}$ & \chemfig{HO-[:30]-[:-30](-[:-90]NH_2)=[:30]NH} \\[0.5cm]
        $v_{34}$ & \chemfig{HO-[:30]-[:-30](-[:-60]NH_2)(-[:-120]HO)-[:30]NH_2} \\
        \bottomrule
    \end{tabular}
    \end{scriptsize}%
    \caption{List of the molecular structures corresponding to the the labeled vertices in the pathways in Figure \ref{fig:pathway1} (the two columns on the left) and Figure \ref{fig:pathway2} (the column on the right).}
    \label{tab:molecular_structure}
\end{table*}

\end{document}